\documentclass[12pt,a4paper]{article}

\usepackage{graphicx}
\usepackage{amsmath}	
\usepackage{amssymb}

\DeclareMathOperator{\atantwo}{atan2}

\usepackage{makecell}
\usepackage{subcaption}

\usepackage{tikz}
\usetikzlibrary{arrows.meta, decorations.pathmorphing}

\usepackage[backend=bibtex,style=numeric,sorting=none,maxbibnames=3]{biblatex}
\usepackage{hyperref}

\begin{document}

\title{Response of Circular Binaries to Monochromatic Gravitational Waves}

\author{Corey McInerney$^{*}$ \and Phil J. Sutton\\[-0.35em]
	\small School of Engineering and Physical Sciences, University of Lincoln,\\[-0.3em]
	\small Brayford Pool, Lincoln LN6 7TS, United Kingdom\\[-0.2em]
	\small $^*$\texttt{cmcinerney@lincoln.ac.uk}}

\maketitle

\begin{abstract}
	
	The exchange in specific orbital energy between a test particle in a Keplerian binary orbit and a continuously incident linearized gravitational wave (GW) is considered. Analytical expressions for the rate of change of energy, semi-major axis and mean motion of the test particle are derived. Numerical solutions illustrate a resonance effect between the GWs frequency and the particle's initial orbital frequency. We distinguish between bounded short-period oscillations and resonant libration. The effect of the GWs angle of incidence with the particle's orbital plane is also considered. Using an illustrative high GW strain we show that changes to eccentricity are bounded while inclination exhibits small linear growth for an initially flat, circular orbit. Long coherence requirements limit the observational significance of GW resonance to astrophysical systems with very short orbital periods.
	
\end{abstract}


\section{Introduction}
\label{Intro}

The exchange of energy between gravitational waves (GW) and the various bodies and particle's which they may interact with as they pervade the universe has been of interest since it was first shown that GWs carry energy \cite{Bondi1957, BoPiRo1959}. Since then the interaction of GWs and binary systems \cite{Mashhoon1978, HuMcYa2013}, dark matter \cite{FlWe2018}, galaxies \cite{Rees1971, Loeb2020} and satellites \cite{YoAh1997} have all been studied. 

In each case, the GW acts as a perturbation to the system and may induce a cumulative resonance effect whereby the response of the system to the incident GW is enhanced \cite{McFoKo2014, ChMaRe2000}. Several methods have been employed to deal with this effect including gravitoelectromagnetism \cite{IoCo2011}, Hamiltonian dynamics \cite{Youssef2017a} and perturbation theory \cite{Mashhoon1978}. This work will focus on the resonant energy exchange between a continuously incident GW and the outer particle of a Keplerian binary system. Previous works have shown the potential for resonance effects when the frequency of the incoming GW, $\omega_{\text{GW}}$, is an integer multiple of the orbiting particle's initial mean motion, $n_{0}$, such that $\omega_{\text{GW}} = mn_{0}$ where $m = 1, 2, 3 \ldots$ \cite{Turner1979, Iv1987, ChMaRe1997}. For the systems like the one of interest here, the absorbed energy from the GW can produce changes in a particle's orbital elements such as semi-major axis, inclination and eccentricity \cite{NeCh1982, Iorio2014, BlJe2022a}. 

The resonant effects induced by GWs on such systems have the potential to amplify orbital perturbations from the GW and open up the range of observable frequencies by which we can detect GWs. Currently, ground-based detectors such as the LIGO, VIRGO and KAGRA interferometers are capable of detecting GWs in the 10 - 10$^{4}$ Hz range \cite{RaBhSa2024}. Sources of these GWs are typically violent, transient events such as compact binary mergers. Space-based detectors like the planned LISA mission, which also uses interferometer technology, will be able to search in the $10^{-4}$ - 1 Hz range where typical sources are thought to be compact binaries and stellar collapse \cite{Gudrun2019}. Yet finding a way of moving even lower in frequency could allow the detection of The Big Bang, primordial fluctuations and inflationary processes \cite{Hughes2003}. 

Opening the range of detectable GW frequency bands allows for a deeper understanding of the processes which shape the universe, enables the testing of current models and the possible discovery of new physics. Considering the types of binary systems in the galaxy, the range of orbital frequencies stretches from 50 Hz for millisecond pulsars \cite{Manchester2017} to $10^{-15}$ Hz for wide stellar binaries \cite{PaSaPa2026}, greatly enhancing the number of sources which could be studied.

In this paper, we derive analytical relations between the energy introduced to a test particle in a circular binary system by a continuously incident GW and the particle's mean motion and semi-major axis. The resulting dynamics are compared to that of a parametrically driven pendulum before the resulting set of coupled differential equations are numerically solved for two resonance conditions ($\omega_{\text{GW}} = n_{0}, 2n_{0}$). The effect that the GWs angle of incidence with the system has on the exchange of energy is also considered as are the libration timescales of several astrophysical systems.

\section{Energy Evolution and Resonance}
\label{Energy_Evol}

\begin{figure}
	\centering
	\begin{tikzpicture}[
		x={(-0.6cm,-0.3cm)}, 
		y={(0.8cm,-0.2cm)}, 
		z={(0cm,1cm)}, 
		>=Stealth
		]
		
		\draw[->, gray!70, dashed, line width=0.6pt] ({2.8*cos(110)}, {2.8*sin(110)},0) -- ({2.8*cos(110) + 4}, {2.8*sin(110)},0) node[left] {$x$};
		\draw[->, gray!70, dashed, line width=0.6pt] ({2.8*cos(110)}, {2.8*sin(110)},0) -- ({2.8*cos(110)}, {2.8*sin(110) + 4},0) node[right] {$y$};
		\draw[->, gray!70, dashed, line width=0.6pt] ({2.8*cos(110)}, {2.8*sin(110)},-4.5) -- ({2.8*cos(110)}, {2.8*sin(110)},0) node[above] {$z$};
		
		\draw[thick, blue!70!black, dashed] plot[domain=0:360, samples=90] ({2.8*cos(\x)}, {2.8*sin(\x)}, 0);
		
		\shade[ball color=cyan!70!black] (0,0,0) circle (0.35cm);
		\node[below left, font=\small] at (0,0,0) {};
		
		\coordinate (particle) at ({2.8*cos(110)}, {2.8*sin(110)}, 0);
		\shade[ball color=orange!90!black] (particle) circle (0.12cm);
		\node[above left, font=\small] at (particle) {};
		
		\draw[->, thick, blue!80!black] plot[domain=110:140, samples=20] ({2.8*cos(\x)}, {2.8*sin(\x)}, 0);
		
		\foreach \xW/\yW in {1/1} {
			\draw[thick, red!80!black, decorate, 
			decoration={snake, amplitude=1.2mm, segment length=4mm, post length=3mm}, ->] 
			({2.8*cos(110)}, {2.8*sin(110)}, -4.5) -- ({2.8*cos(110)}, {2.8*sin(110)}, -0.1);
		}
		
		\node[red!90!black, right, align=left, font=\small] at ({2.8*cos(110)-5}, {2.8*sin(110)-5}, -4.5) {
			GW\\
		};
		
	\end{tikzpicture}
	\caption{A GW (red wave) incident upon a test particle (orange sphere) from the bottom to top along the $z$-direction. The test particle is orbiting a larger central mass (green sphere).}
	\label{Diagram}
\end{figure}
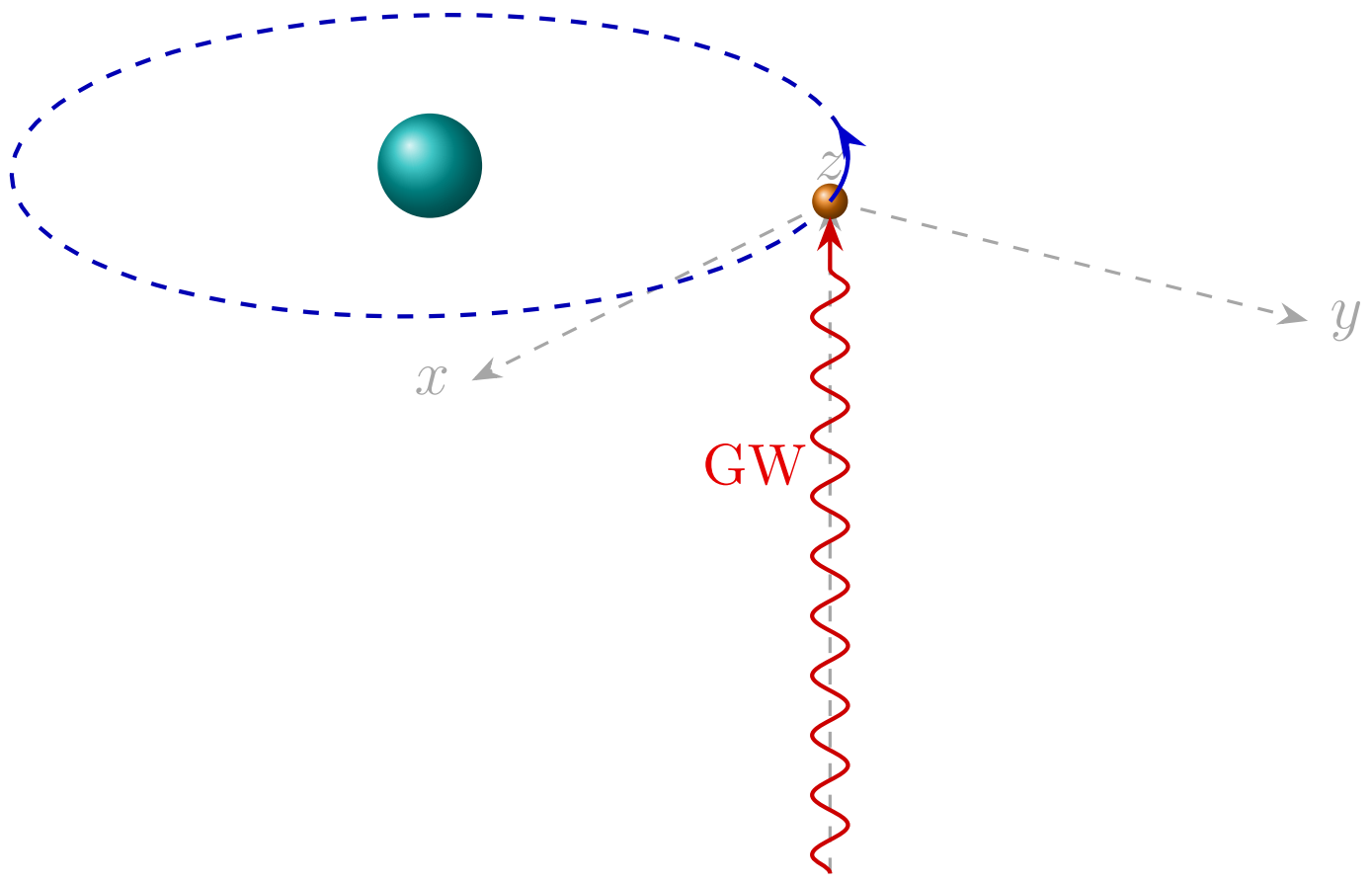

In this work we are interested in a Keplerian binary analogous to a star-planet, planet-moon or planetary ring system where a test particle orbits around a much larger central mass. We work in the linearized theory of general relativity and consider a plane, monochromatic GW defined in the transverse-traceless gauge and weak-field limit where the metric perturbation due to a GW incident along the test particle's $z$-direction is

\begin{gather}
	h_{ij}(t, z) = \begin{pmatrix}
		h_{+}\cos(kz - \omega_{\text{GW}}t + \phi_{+}) & h_{\times}\cos(kz - \omega_{\text{GW}}t + \phi_{\times}) \\
		h_{\times}\cos(kz - \omega_{\text{GW}}t + \phi_{\times}) & -h_{+}\cos(kz - \omega_{\text{GW}}t + \phi_{+})
	\end{pmatrix}
	\label{TTSol}
\end{gather}

Equation (\ref{TTSol}) describes a GW with two transverse modes of polarization, commonly referred to as plus (+) and cross ($\times$). Each of these polarizations has their own independent amplitude ($h_{+}$ and $h_{\times}$ respectively). $\phi_{+, \times}$ represents the phase of each polarization, $\omega_{\text{GW}}$ is the frequency of the GW, $k$ the wave number and $t$ is time.

A graphic illustrating the binary and incident GW is given in Figure \ref{Diagram}. 

When a GW is incident upon the ring, it will induce an acceleration in both the $x$ and $y$ directions where the acceleration components are given by \cite{MTW}

\begin{equation}
	A_{x} = \frac{1}{2}(\ddot{h}_{+}x + \ddot{h}_{\times}y)
	\label{Ax}
\end{equation}

\begin{equation}
	A_{y} = \frac{1}{2}(\ddot{h}_{\times}x - \ddot{h}_{+}y)
	\label{Ay}
\end{equation}

with overdots referring to differentiation with respect to time. 

The energy imparted upon the particle can be thought of as the work done by the incident GW. Working in natural units ($G = c = 1$), in the particle's center-of-mass frame the energy of the system can be rewritten using Kepler's third law such that \cite{ShGiXu2026}

\begin{equation}
	\mathcal{E} = \frac{1}{2}\mu_r|\dot r^2| - \frac{\mu_r\mu_g}{r} = -\frac{\mu_r\mu_g}{2a}
\end{equation}

where $\mu_r$ is the systems reduced mass, $r$ is the length of the separation vector (radius), $\mu_g$ is the systems gravitational parameter, $a$ is the orbiting particle's semi-major axis and $n$ is its mean motion. 

It is convenient to introduce the specific orbital energy, $(E=\mathcal{E}/\mu_r)$, which is independent of the test-particle mass. Using Kepler's third law, $(n^2a^3=\mu_g)$, gives

\begin{equation}
	E = -\frac{\mu_g}{2a} = -\frac{\mu_g^{2/3}n^{2/3}}{2}
	\label{E_Init}
\end{equation}

We can get the change in specific energy of the particle over time caused by the GW by considering the induced accelerations in Equation (\ref{Ax}) and (\ref{Ay}) \cite{HuMcYa2013}

\begin{equation}
	\dot{E} = (A_{x}\dot{x} + A_{y}\dot{y}) = \frac{1}{2}\left(\ddot{h}_{+}(x\dot{x} - y\dot{y}) + \ddot{h}_{\times}(x\dot{y} + y\dot{x})\right)
	\label{E_dot}
\end{equation}

This shows that the energy introduced into the system by the GW evolves linearly with the strain amplitude and is dependent upon the GW phase. Additionally, the GW is coupled to the orbital motion of the particle. We assume that the GW is continuously incident upon the system with a constant frequency and amplitude. At this stage the GW source is not of significant interest.

GWs which have the two polarization modes $90^{\circ}$ out of phase with each other, but have the same amplitude ($h_{+} = h_{\times} = h_{0}$) are called circularly polarized waves \cite{Maggiore2007}. As such, using Equation (\ref{TTSol}) we can describe the $+$ and $\times$ mode contribution to the GW as

\begin{equation}
	h_{+}(t) = h_{0}\cos(\omega_{\text{GW}}t-kz)
\end{equation}

\begin{equation}
	h_{\times}(t) = h_{0}\sin(\omega_{\text{GW}}t-kz)
\end{equation}

Taking the second time derivative of each mode gives

\begin{equation}
	\ddot{h}_{+} = -\omega_{\text{GW}}^2h_{0}\cos(\omega_{\text{GW}}t)
	\label{ddot_+}
\end{equation}

\begin{equation}
	\ddot{h}_{\times} = -\omega_{\text{GW}}^2h_{0}\sin(\omega_{\text{GW}}t)
	\label{ddot_cross}
\end{equation}

\section{Resonance and the Pendulum Equation}
\label{R&P}

We restrict ourselves to working with a circular orbit where the $(X, Y)$ coordinates of the particle are described by

\begin{equation}
	X = a\cos(nt)
	\label{x}
\end{equation}

\begin{equation}
	Y = a\sin(nt)
\end{equation}

The derivatives of the above are given by 

\begin{equation}
	\dot{X} = -an\sin(nt)
	\label{dot_x}
\end{equation}

\begin{equation}
	\dot{Y} = an\cos(nt)
	\label{dot_y}
\end{equation}

Following \cite{HuMcYa2013}, we consider the GW to be defined in a separate frame $(x, y, z)$ to that of the orbiting particle $(X, Y, Z)$, with the two frames related by a series of Euler rotations such that

\begin{equation}
	\begin{bmatrix}
		x \\
		y \\
		z
	\end{bmatrix} = a 
	\begin{bmatrix}
		\cos(nt - \phi)\cos\theta \\
		\sin(nt - \phi) \\
		\cos(nt - \phi)\sin\theta
	\end{bmatrix}
\end{equation}

where the inclination angle between the GWs direction of propagation and the particle's orbital plane is $\theta$ and $\phi$ is a constant phase offset defining the relative orientation. Specifically, $\theta = \cos^{-1}(\hat{\textbf{k}}\cdot\hat{\textbf{L}})$ where $\hat{\textbf{k}}$ is the GW propagation direction and $\hat{\mathbf L}$ is the orbital angular-momentum unit vector.

Substituting Equation (\ref{ddot_+}) - (\ref{dot_y}) into Equation (\ref{E_dot}) and working through the trigonometry (see Appendix \ref{Energy}) produces

\begin{equation}
	\dot{E} = \frac{ a^2n\omega_{\text{GW}}^2}{2}[C_{+}\sin((\omega_{\text{GW}} + 2n)t-2\phi) + C_{-}\sin((\omega_{\text{GW}} -2n)t + 2\phi)]
	\label{E_dot_final}
\end{equation}

where we define two dimensionless amplitude coefficients

\begin{equation}
	C_{+} = \frac{1}{4}h_{0}(1-\cos(\theta))^2
	\label{C+}
\end{equation}

and 

\begin{equation}
	C_{-} = -\frac{1}{4}h_{0}(1+\cos(\theta))^2
	\label{C-}
\end{equation}

Each coefficient is dependent upon the GWs amplitude and inclination angle, $\theta$, between the GW and $z$-axis. This derivation is valid for a circular orbit, where the mean anomaly is $\mathcal{M}=nt$. 

Equation (\ref{E_dot_final}) shows that the energy input into the ring particle's orbit by the GW oscillates via a coupling between the GWs frequency and the orbital frequency of the particle. This provides a condition for resonance, $\omega_{\text{GW}} = 2n$, as when this occurs, the frequency related parts of the bracketed term in Equation (\ref{E_dot_final}) is zero and the sine terms lose their time dependence, becoming linear. 

Upon application of the above resonance condition, the arguments of the two sine terms in Equation (\ref{E_dot_final}) leads to fast ($\Phi = (\omega_{\text{GW}} + 2n)t - 2\phi$) and slow ($\Delta = (\omega_{\text{GW}} - 2n)t + 2\phi$) oscillations. The fast term contributes bounded oscillations with amplitude $\frac{\omega_{\text{GW}}^2C_+}{[n_0(\omega_{\text{GW}}+2n_0)]}$ and averages out over many successive orbital periods. The slow resonant governs the long-term evolution of the system, leaving

\begin{equation}
	\dot{E} = \frac{a^2n\omega_{\text{GW}}^2}{2}[C_{-}\sin(\Delta)]
	\label{E_average}
\end{equation}

As $n$ is a function of time, mathematically it would be more rigorous to define the slow term as an integral where $\Delta(t) = \int(\omega_{\text{GW}} -2n(t))dt + 2\phi$. The following analysis is valid assuming that forcing from the GW is weak and orbital evolution is adiabatic. Specifically, we assume that the orbital elements evolve on a timescale much longer than the orbital period such that

\begin{equation}
	\left|\frac{\dot n}{n^2}\right| \ll 1,
	\qquad
	\left|\frac{\dot a}{an}\right| \ll 1
\end{equation}

allowing the orbital motion to be treated as approximately periodic over a single orbit. Under this separation of timescales, the rapidly oscillating terms average to zero over many orbital periods while the slowly varying resonant contribution accumulates secularly. The ultra-low-frequency limit $\omega_{\text{GW}}\ll n$ of \cite{Iorio2014} is distinct from the $2n$ resonant regime studied here in that adiabatic non-resonant regime the averaged semi-major axis is unchanged while other orbital elements can vary. In contrast, near $2n$ the resonant part of the tidal work need not average to zero. The methods are complementary, not competing general solutions.

Invoking Kepler's third law, it follows that $\frac{\dot{n}}{n}=\frac{3}{2}\frac{\dot{E}}{E}$ and so

\begin{equation}
	\dot{n} = -\frac{3}{2}\omega_{\text{GW}}^{2}C_-\sin(\Delta)
	\label{n_ratio}
\end{equation}

Thus considering $\ddot{\Delta} = -2\dot{n}$, Equation \eqref{n_ratio} leads to 

\begin{equation}
	\ddot{\Delta} - 3\omega_{\text{GW}}^2|C_-|\sin(\Delta) = 0
	\label{PendulumEq}
\end{equation}

which is analogous to that describing the motion of a pendulum of amplitude $3\omega_{\text{GW}}^2C_{-}$ (details of similar derivations for pure $+$, pure $\times$ and mixed polarization GWs are given in Appendix \ref{GVEs}). Thus, the resonant oscillation of the orbiting particle is dependent only upon the properties of the incident GW (frequency, amplitude and angle of incidence). The pendulum approximation then tells us that the GW strain amplitude scales with the system. Indeed, a GW amplitude is a dimensionless fractional strain applied upon the binary whereby the physical displacement it causes is strictly proportional to the size of the system it is passing through. A binary with a larger semi-major axis will experience a larger absolute physical displacement, but any fractional displacement, $\delta a/a$, will remain identical to that of a smaller binary.

Furthermore, Equation (\ref{PendulumEq}) is completely scale-invariant. The exact same mathematical model applies universally across vastly different astrophysical systems. The resonance dynamics are mathematically identical as long as the orbital period is tuned to the GW frequency, the system will behave exactly according to the pendulum equation.

\begin{figure}
	\centering\includegraphics[width=.82\linewidth]{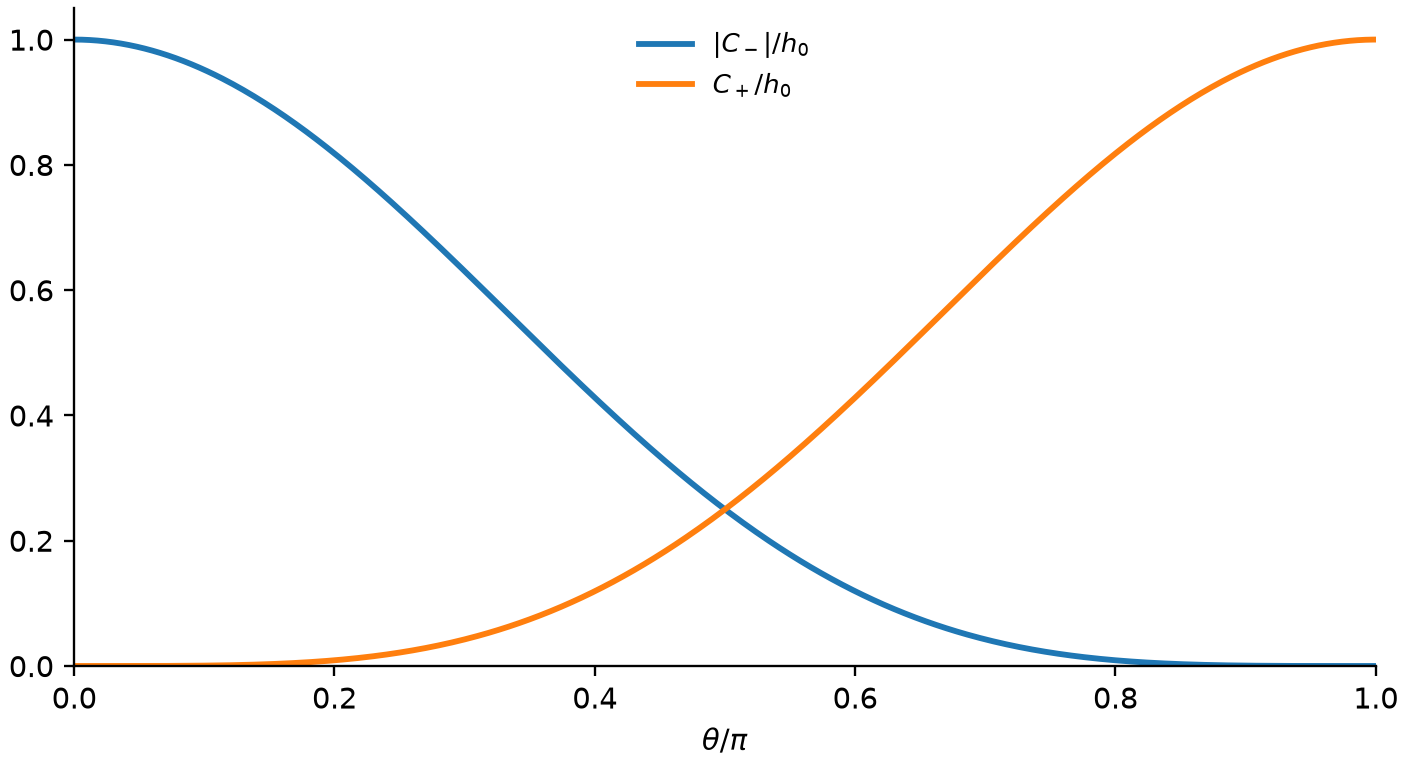}
	\caption{Angular coupling for the adopted circular-polarization handedness. At $\theta=\pi$ the resonant coefficient vanishes while at $\theta=\pi/2$, $|C_\pm|=h_0/4$. Reversing helicity exchanges the two curves.}
	\label{fig:hel}
\end{figure}

The angular factor in the $C_{\pm}$ coefficients plays a decisive role in the dynamics of our system. Figure \ref{fig:hel} plots the coupling of the coefficients where we see that at $\theta=0$, $C_-=-h_0$ and $C_+=0$. At $\theta=\pi/2$, $C_-=-h_0/4$ and $C_+=h_0/4$, yet when $\theta=\pi$, $C_-=0$ while $C_+=h_0$. Hence the resonance coefficient vanishes identically at $\theta=\pi$ for the adopted handedness. There is no $\theta\leftrightarrow\pi-\theta$ symmetry for a GW of circular polarization.

The libration frequency, $\omega_{\text{lib}}$, is the square root of the amplitude in Equation (\ref{PendulumEq}). Applying the resonance condition $\omega_{\text{GW}} = 2n$ we have

\begin{equation}
	\omega_{\text{lib}} = 2n\sqrt{3|C_{-}|}
	\label{Omega_lib}
\end{equation}

As the period of libration is $T_{\text{lib}} = \frac{2\pi}{\omega_{\text{lib}}}$, when compared with $T_{\text{orb}}$ this leads to the following expression for the libration period \cite{MuDe1999}

\begin{equation}
	T_{\text{lib}} = \frac{T_{\text{orb}}}{2\sqrt{3|C_{-}|}}
	\label{T_lib}
\end{equation}

Thus, for a GW of amplitude $h_{0} = 10^{-15}$, $\theta = \frac{\pi}{4}$, the period of libration is $T_{\text{lib}} = 1.07\times 10^{7} T_{\text{orb}}$. 

The libration timescale has wider implications for the resonant detection of GWs from binary systems, providing constraints on the astrophysical sources where this effect can be observed. Considering a planetary ring particle such as those orbiting Saturn where orbital periods are on the order of $5-90$ hours for the inner D and outer E rings respectively, this leads to $T_{\text{lib}} \approx 6.1\times 10^{3} - 1.1\times 10^{5}$ years making resonant accumulation extremely slow and secular trapping impossible over observational timescales for this system.

Other orbital systems which have been considered as having potential resonances with GWs include planetary systems \cite{IsSa2011, Youssef2018b}, galaxies \cite{FuMa1979} and pulsars \cite{BlJe2022a}. Libration timescales for these systems are on the order of $2.57\times 10^{6} - 1.77\times 10^{9}$ years, $2.44\times10^{14}$ years, and $1.08\times 10^{3}$ years respectively for Solar System planets ($T_{\text{orb}} \simeq 0.24 - 165.42$ years) \cite{Bevelacqua}, stars in the solar neighborhood ($T_{\text{orb}} \simeq 2.28\times10^7$ years) \cite{MrUdSk2019} and pulsars ($T_{\text{orb}} \simeq 53$ minutes) \cite{PaLuJi2023}. These numbers show that the libration timescale for planetary systems renders the effect dynamically negligible compared to secular perturbations from other bodies. $T_{\text{lib}}$ for galactic systems exceeds the age of the universe which makes these systems observationally irrelevant for this type of measurement. 

However, more promising environments may include compact binaries and pulsar systems where orbital periods can be minutes to hours and where the surrounding gravitational environment is significantly stronger. In such systems the libration timescale may fall within astrophysically relevant intervals, potentially allowing resonant GW-induced perturbations to accumulate secularly over observable timescales. The NANOGrav collaboration, which uses radio waves to observe pulsars have public datasets which currently span 15 years \cite{NANOGrav15} and data from the Parkes Observatory dates back as far as the 1990s \cite{Parkes}. Yet neither of these datasets reaches the above $O(10^3)$ years estimate of pulsar libration periods. 

If the full expression for the rate of change of energy Equation (\ref{E_dot_final}) is substituted into Equation (\ref{PendulumEq}) a fuller description of the pendulum-like nature of the system is revealed where Equation (\ref{PendulumEq}) becomes

\begin{equation}
	\ddot{\Delta} - 3\omega_{\text{GW}}^{2}[C_{-}\sin(\Delta) + C_{+}\sin(\Phi)] = 0
	\label{Pendulum_full}
\end{equation}

where we have defined the fast oscillation to be $\Phi = (\omega_{\text{GW}} + 2n)t - 2\phi$. 

Equation (\ref{Pendulum_full}) is analogous to that of a parametrically driven pendulum where the right hand side acts as a high frequency forcing term. 

While Equation (\ref{PendulumEq}) describes an integrable system with a bounded libration trap, the re-introduction of the fast circulating phase in Equation (\ref{Pendulum_full}) subjects the system to high-frequency periodic forcing \cite{Butikov2002}. Fast oscillations do not create a resonance trap as the second time derivative of this phase is $\ddot{\Phi} = 2\dot{n}$, which is not necessarily positive since $n$ can increase or decrease as energy is exchanged with the incident GW. Nevertheless, for positive GW frequency and prograde orbital motion, $\dot{\Phi}=\omega_{\mathrm{GW}}+2n>0$, ensuring that the fast phase continues to circulate rather than becoming stationary. Consequently, unlike the slow phase, which can undergo resonant trapping when $\omega_{\mathrm{GW}}\simeq2n$, the fast component produces bounded, non-resonant oscillations that average out over sufficiently long timescales within the weak-forcing approximation. 

The inclusion of this non-resonant component modifies the phase-space structure of the averaged pendulum. In near-integrable Hamiltonian systems, periodic perturbations can deform invariant trajectories and alter the structure of the separatrix, potentially producing separatrix splitting and localized chaotic behaviour \cite{Morbidelli2002}. However, such behaviour is not guaranteed solely by the presence of high-frequency forcing.

\begin{figure}
	\centering
	\subfloat[$h_{0}=10^{-2}$]{\label{Large}\includegraphics[width=0.5\textwidth]{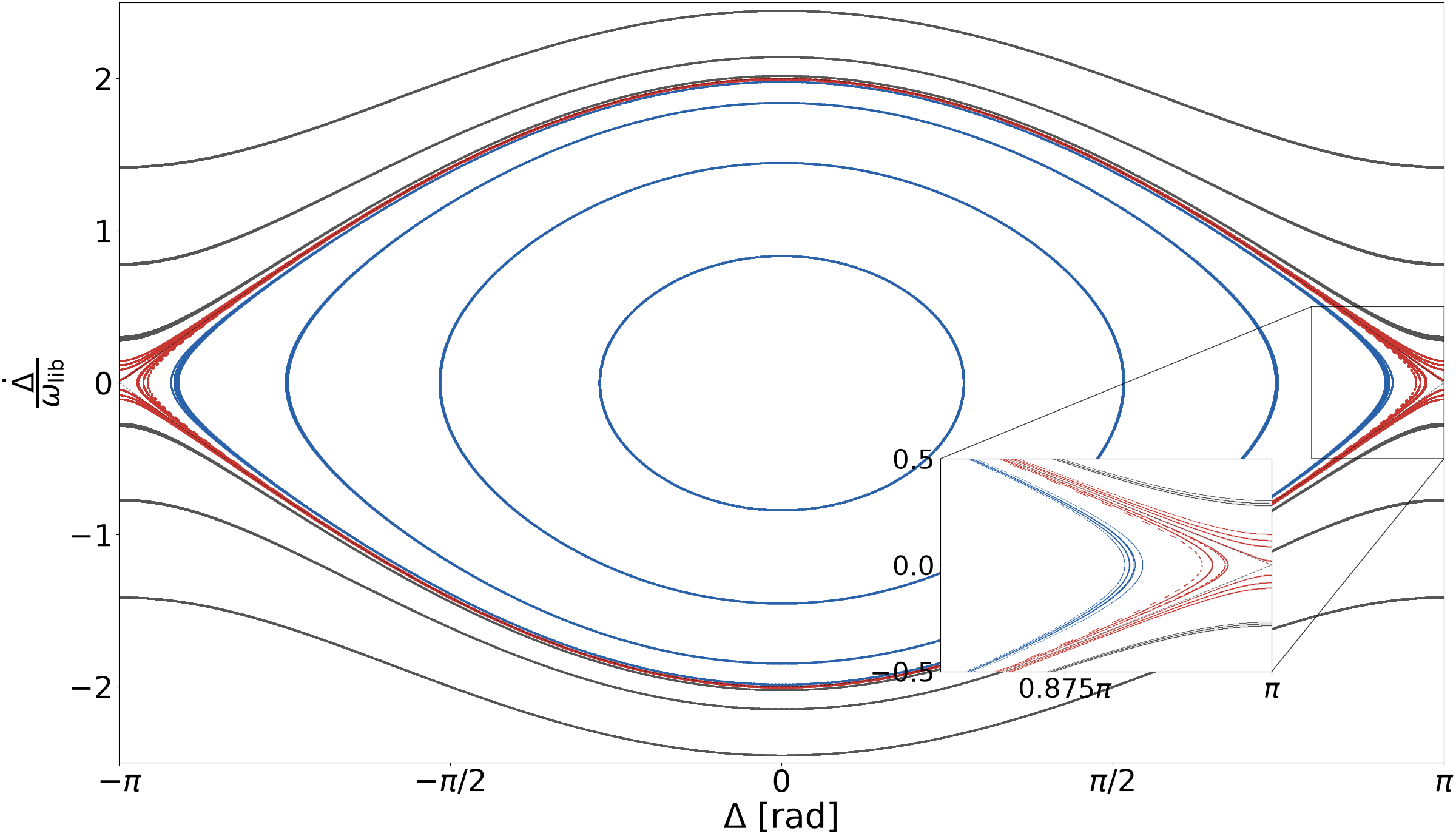}}\hfill
	\subfloat[$h_{0}=10^{-7}$]{\label{Small}\includegraphics[width=0.5\textwidth]{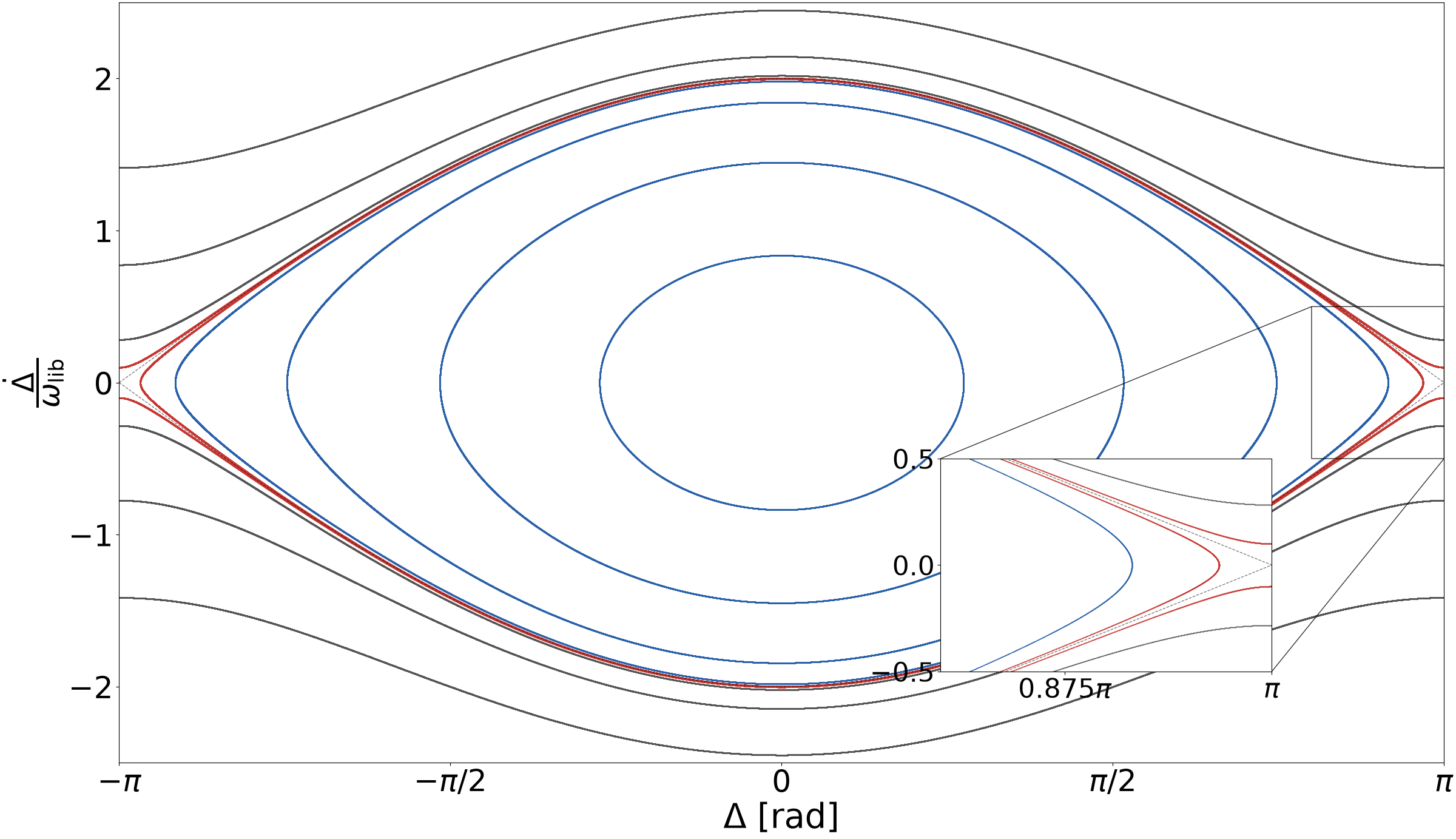}}
	\caption{Surface of section for the full fast and slow oscillation pendulum equation. Blue trajectories originate within the unforced	libration region, grey trajectories within the circulating region, and red trajectories near the unforced separatrix. The dashed curve denotes the unforced separatrix and is not an invariant boundary of the driven	system. We have used $\theta=\pi/4$ and $\omega_{\mathrm{GW}}=2n_0$ for each plot.}
	\label{fig:KAM}
\end{figure}

To investigate the stability of the resonance structure under the influence of the non-resonant oscillatory component, the full reduced pendulum equation was integrated numerically and analyzed using stroboscopic Poincar{\'e} sections. For the adopted circular polarization, the equation of motion takes the form

\begin{equation}
	\ddot{\Delta}+\omega_{\mathrm{lib}}^2\sin(\Delta)
	=3\omega_{\mathrm{GW}}^2C_+\sin(\Phi)
\end{equation}

where $\omega_{\mathrm{lib}}^2=3\omega_{\mathrm{GW}}^2|C_-|$. The slow and fast phases satisfy $\dot{\Delta}=\omega_{\mathrm{GW}}-2n$ and $\dot{\Phi}=\omega_{\mathrm{GW}}+2n$, respectively. Consequently, the external driving phase can be expressed as $\chi=\Phi+\Delta=2\omega_{\mathrm{GW}}t+\chi_0$. The Poincar{\'e} sections were constructed by recording the phase-space coordinates $(\Delta,\dot{\Delta}/\omega_{\mathrm{lib}})$ whenever $\chi=\chi_0\pmod{2\pi}$, ensuring that successive points correspond to the same phase of the external periodic forcing.

Figure \ref{fig:KAM} presents the resulting sections for two GW strain amplitudes, $h_0=10^{-7}$ and $h_0=10^{-2}$, with incidence angle $\theta=\pi/4$ and $\omega_{\mathrm{GW}}=2n_0$. The stronger-strain case is included as an illustrative numerical stress test of the reduced model rather than a physically representative astrophysical GW amplitude. The trajectories are classified according to their initial energy in the unforced pendulum of the form

\begin{equation}
	H_0=\frac{\dot{\Delta}^2}{2}+\omega_{\mathrm{lib}}^2(1-\cos(\Delta))
\end{equation}

for which the separatrix occurs at $H_{0,\mathrm{sep}}=2\omega_{\mathrm{lib}}^2$. Blue trajectories are initialized within the unforced libration region, grey trajectories within the circulating region, and red trajectories close to the separatrix. These classifications refer to the initial conditions; $H_0$ is not conserved when the non-resonant forcing is included.

In the weak-strain case, the phase-space structure is predominantly regular, with smooth nested trajectories around the stable resonant equilibrium and circulating trajectories outside the libration region. Trajectories initialized close to the separatrix also remain organized at the resolution of the numerical section. This behavior is consistent with the weak influence of the rapidly oscillating component on the underlying resonant dynamics.

For the larger strain amplitude, the central libration region remains well defined, while trajectories near the unperturbed separatrix exhibit more pronounced deformation and separation, particularly in the vicinity of the hyperbolic points at \(\Delta=\pm\pi\). The enlarged region of Figure 4 illustrates this local change in phase-space structure. Nevertheless, the resulting trajectories remain largely organized, and the sections do not reveal a clearly resolved, broad stochastic layer. The observed deformation therefore provides evidence that the non-resonant forcing modifies the resonance boundary, but does not by itself demonstrate chaotic diffusion or long-term stochastic transport.

The persistence of regular-looking trajectories is qualitatively compatible with near-integrable Hamiltonian dynamics. However, establishing the existence of chaotic layers would require additional diagnostics, such as calculations of separatrix splitting, Lyapunov exponents or systematic orbital diffusion. Such behavior has previously been investigated in more general GW-perturbed Keplerian systems (see \cite{ChMaRe1996a, ChMaRe1999b}). The present calculation instead illustrates how the non-resonant oscillatory component modifies the phase-space structure of the reduced circular-orbit resonance model, while leaving the central libration region comparatively regular for the parameters examined.

\section{Orbital Evolution}
\label{OrbEvol}

\begin{figure}
	\centering
	\subfloat[Semi-major Axis]{\label{a}\includegraphics[height=0.3\textheight]{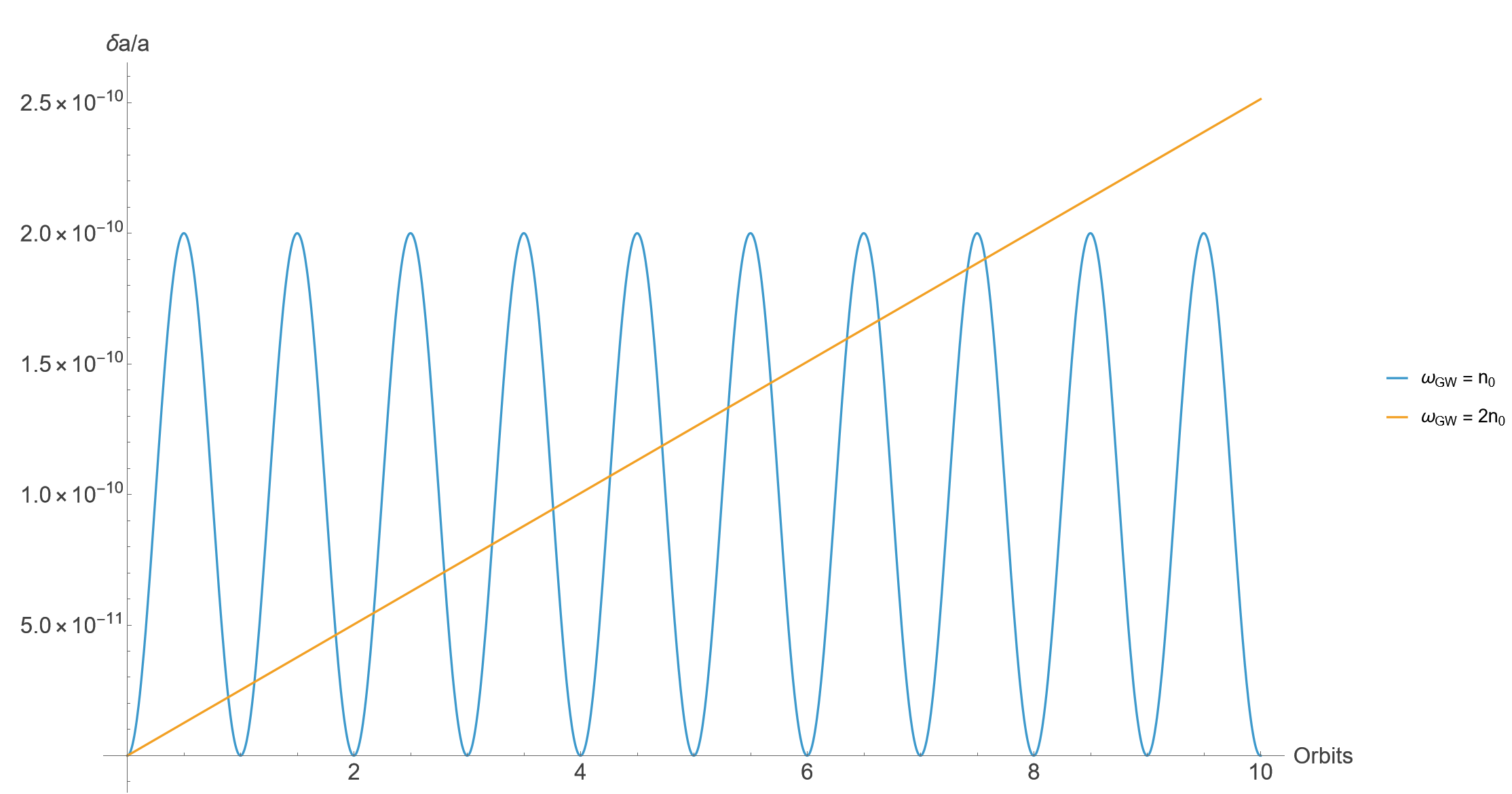}}\hfill
	\subfloat[Specific Energy]{\label{b}\includegraphics[height=0.3\textheight]{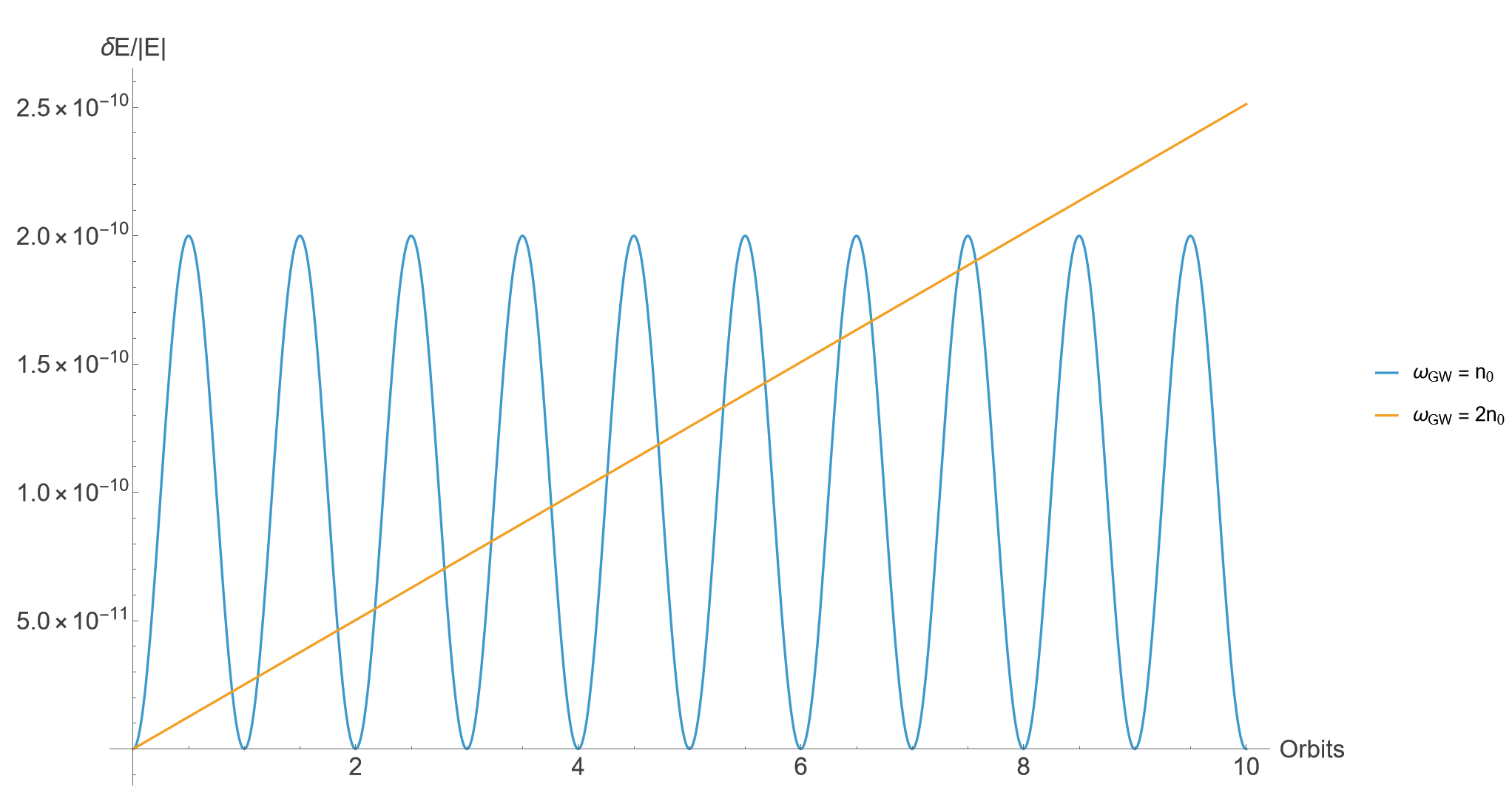}}\hfill
	\subfloat[Mean motion]{\label{c}\includegraphics[height=0.3\textheight]{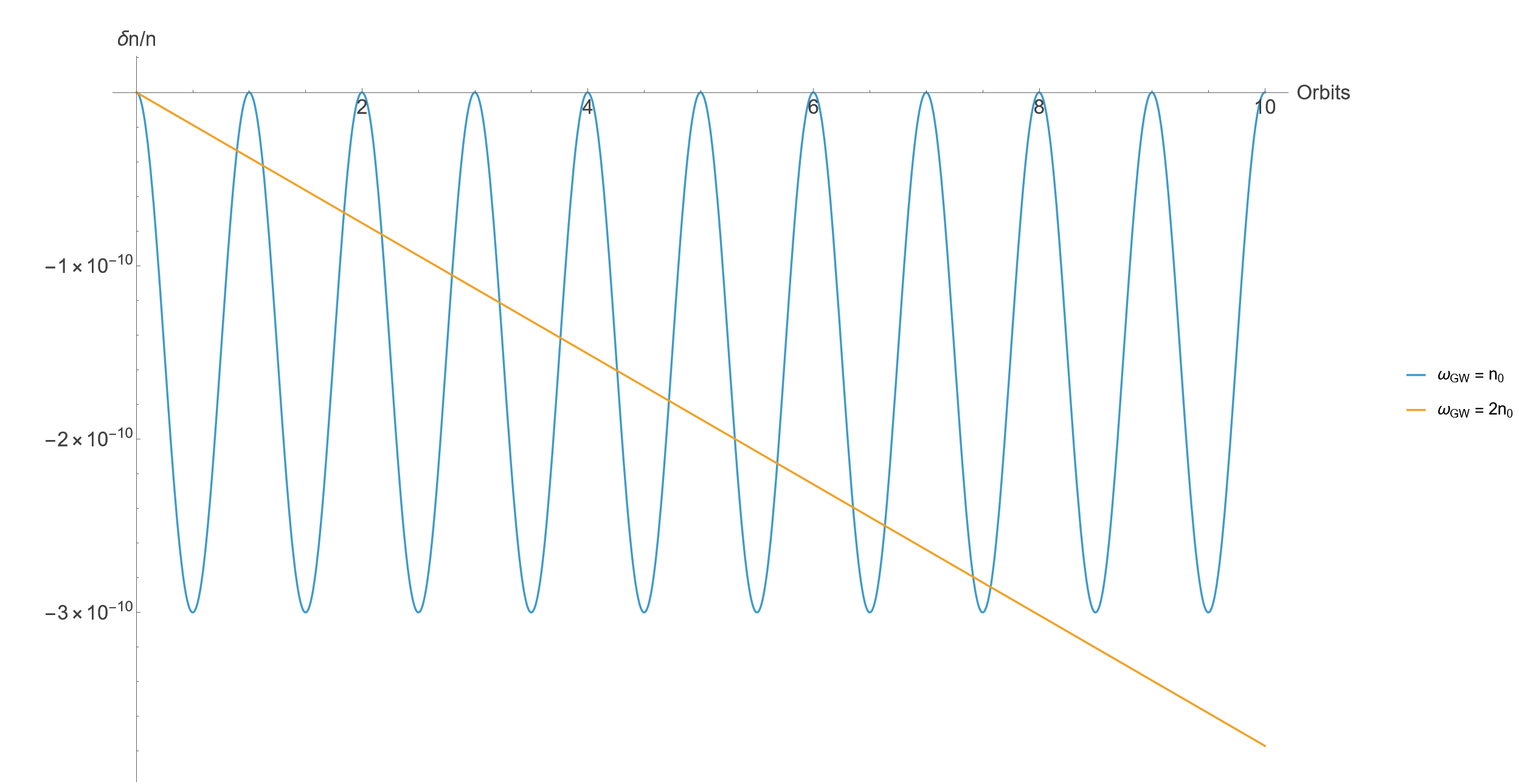}}
	\caption{Relative change in semi-major axis, specific energy and mean motion over 10 orbits. Initial conditions are $h_{0}=10^{-10}$, $a=1$, $\theta=0$ and $\Delta=-0.01$. The GW strain amplitude has been exaggerated for visual clarity.}
	\label{fig:10}
\end{figure}

By again considering that from Kepler's third law, $E \propto -\frac{1}{a}$, and in a similar vein to the above derivation of Equation (\ref{n_ratio}), we can construct an equation for the rate of change of semi-major axis.

\begin{equation}
	\dot{a} = \frac{2a^2\dot{E}}{\mu_g} = \frac{a\omega_{\text{GW}}^2C_-\sin(\Delta)}{n(a)}
	\label{a_dot}
\end{equation}

Together, Equation (\ref{E_dot_final}), (\ref{n_ratio}) and (\ref{a_dot}) represent a coupled system of differential equations which describe the evolution of an orbiting particle's total energy, mean motion and semi-major axis due to the continuous passage of a GW. By taking the initial energy of the system to be given by Equation (\ref{E_Init}), we use the \verb|NDSolve| function provided by MATHEMATICA to solve the above system of equations for the conditions $\omega_{\text{GW}} = n_{0}$ and $\omega_{\text{GW}} = 2n_{0}$. The relative change in semi-major axis, energy and mean motion from this analysis are plotted in Figure \ref{fig:10}. 

The oscillatory nature of the deviations to $a$, $E$ and $n$ are immediately evident from Figure \ref{fig:10}. As the particle progresses along its orbit, the relative phase of the GW changes. When $\omega_{\text{GW}} = n_{0}$, the particle is off resonance with the GW and circulates. When the two are in phase, the GW adds positive energy to the system, moving the particle to a wider orbit and increasing the semi-major axis. Consequently, the mean motion must decrease if $a$ increases. This is reflected in the plots. 

While on resonance ($\omega_{\text{GW}} = 2n_{0}$), the slow phase satisfies $\dot{\Delta} = \omega_{\text{GW}} - 2n \approx 0$ and so over a short time window $\Delta$ is essentially frozen and $\dot E$ is constant, giving the linear drift seen in the plots. More precisely Figure \ref{fig:10} uses initial conditions of $\theta=0$ and $\Delta = -0.01$, therefore $\dot{E}>0$ and the change in energy is positive. Setting $\Delta(0) = 0$ would cause the change in specific energy to disappear at first order while starting with $\Delta>0$ would cause $\dot{E}<0$ over short timescales. Additionally, for $\theta=0$, there is no contribution from the fast term ($C_+=0$).

\begin{figure*}
	
	\begin{subfigure}{.475\linewidth}
		\includegraphics[width=\linewidth]{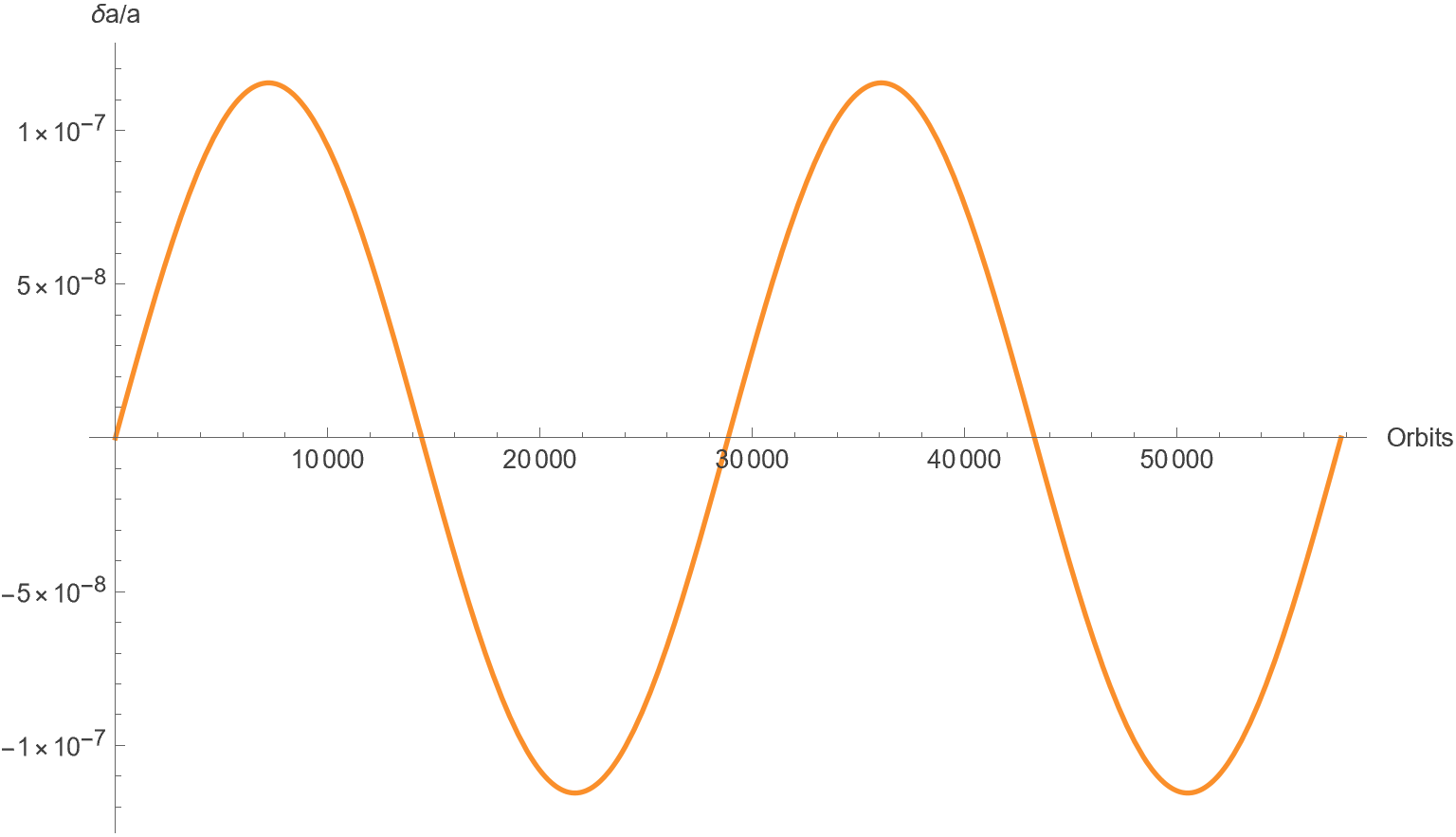}
		\caption{Semi-major Axis}
		\label{a_long}
	\end{subfigure}
	\begin{subfigure}{.475\linewidth}
		\includegraphics[width=\linewidth]{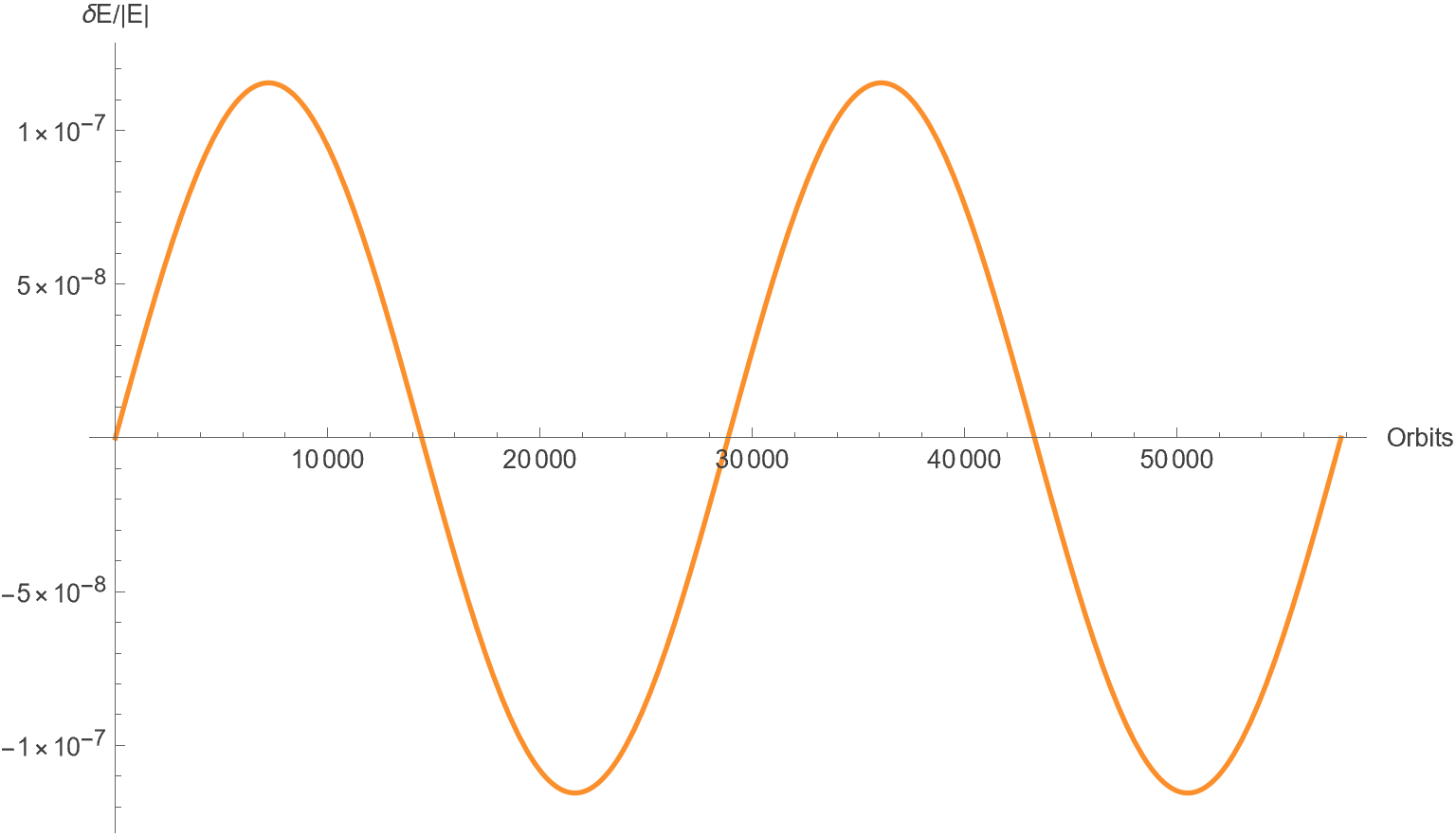}
		\caption{Specific Energy}
		\label{E_long}
	\end{subfigure}
	
	\begin{subfigure}{.475\linewidth}
		\includegraphics[width=\linewidth]{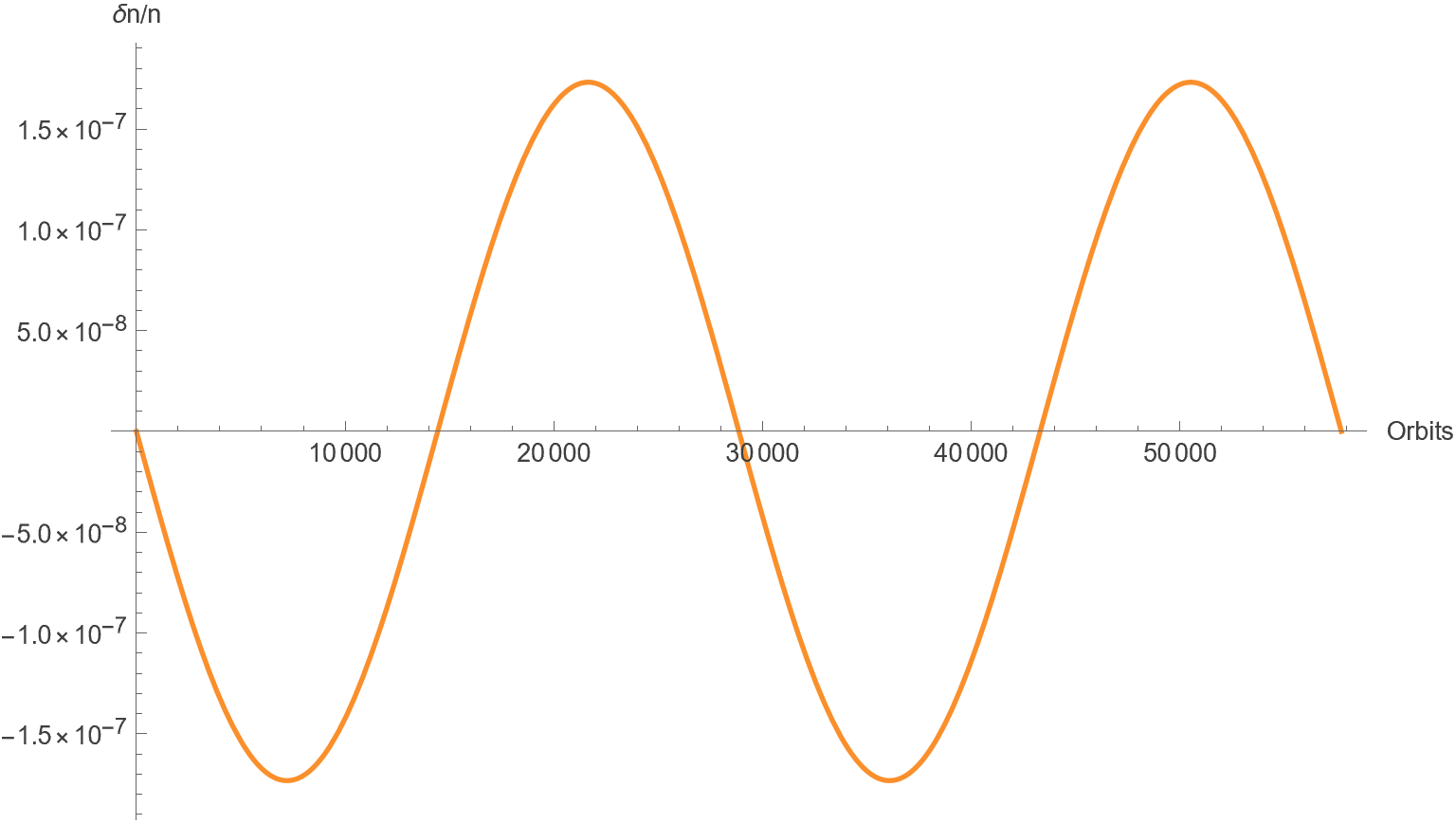}
		\caption{Mean Motion}
		\label{n_long}
	\end{subfigure}
	\begin{subfigure}{.475\linewidth}
		\includegraphics[width=\linewidth]{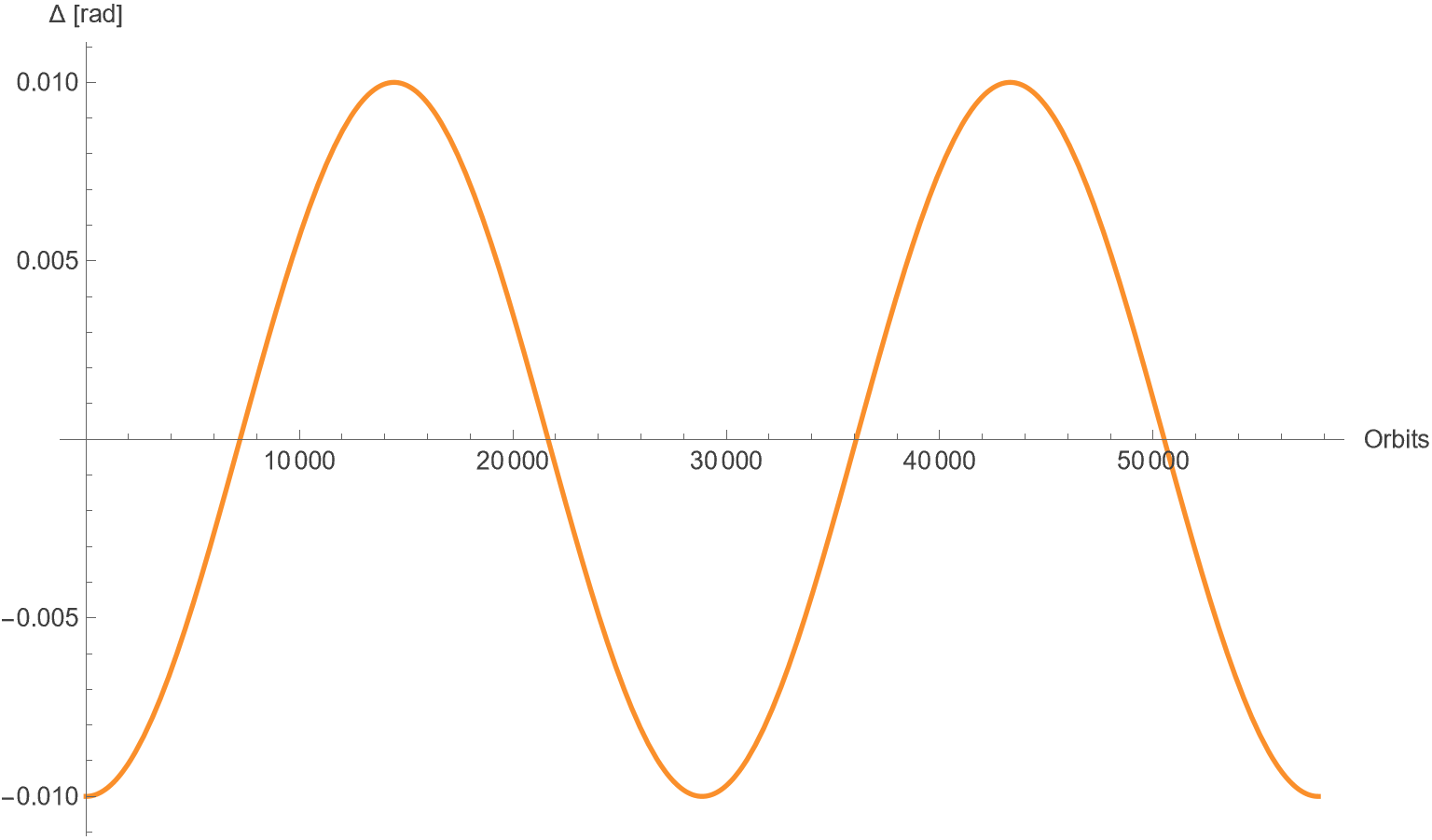}
		\caption{Slow Phase, $\Delta$}
		\label{Delta_long}
	\end{subfigure}
	
	\caption{Relative change in semi-major axis, specific energy and mean motion as well as the evolution of the slow resonant phase over two libration periods. Initial conditions are $\omega_{\text{GW}} = 2n$, $h_{0}=10^{-10}$, $a=1$, $\theta=0$ and $\Delta=-0.01$. The GW strain amplitude has been exaggerated for visual clarity.}
	\label{fig:Long}
\end{figure*}

When on resonance, the slow phase and orbital elements can be seen to oscillate at the timescale predicted by Equation \eqref{T_lib} in Figure \ref{fig:Long}. $\Delta$ is bounded around a stable equilibrium of $\Delta = 0$, indicative of libration. 

\begin{figure}
	\centering
	\subfloat[Semi-major Axis]{\label{theta_a}\includegraphics[height=0.3\textheight]{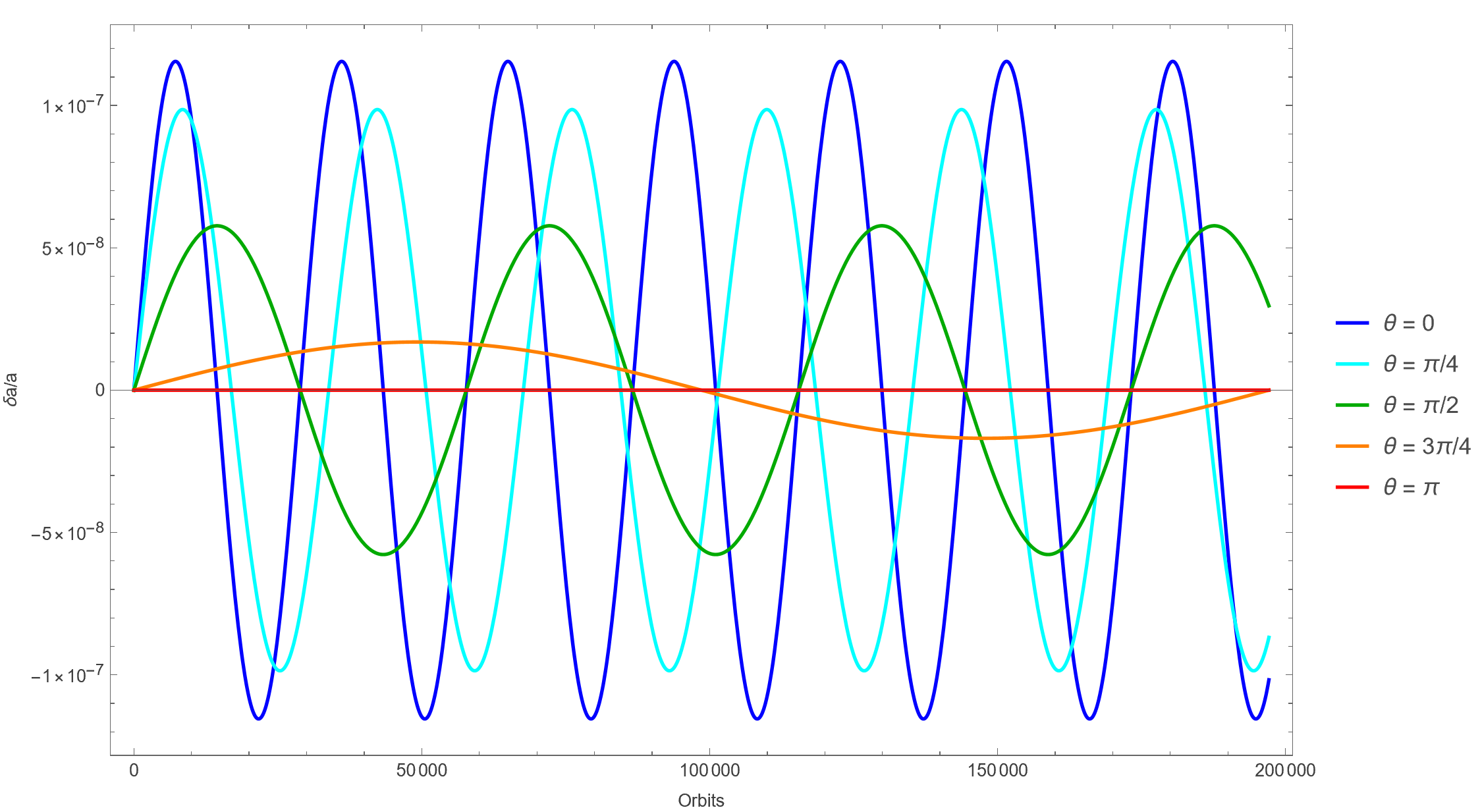}}\hfill
	\subfloat[Energy]{\label{theta_b}\includegraphics[height=0.3\textheight]{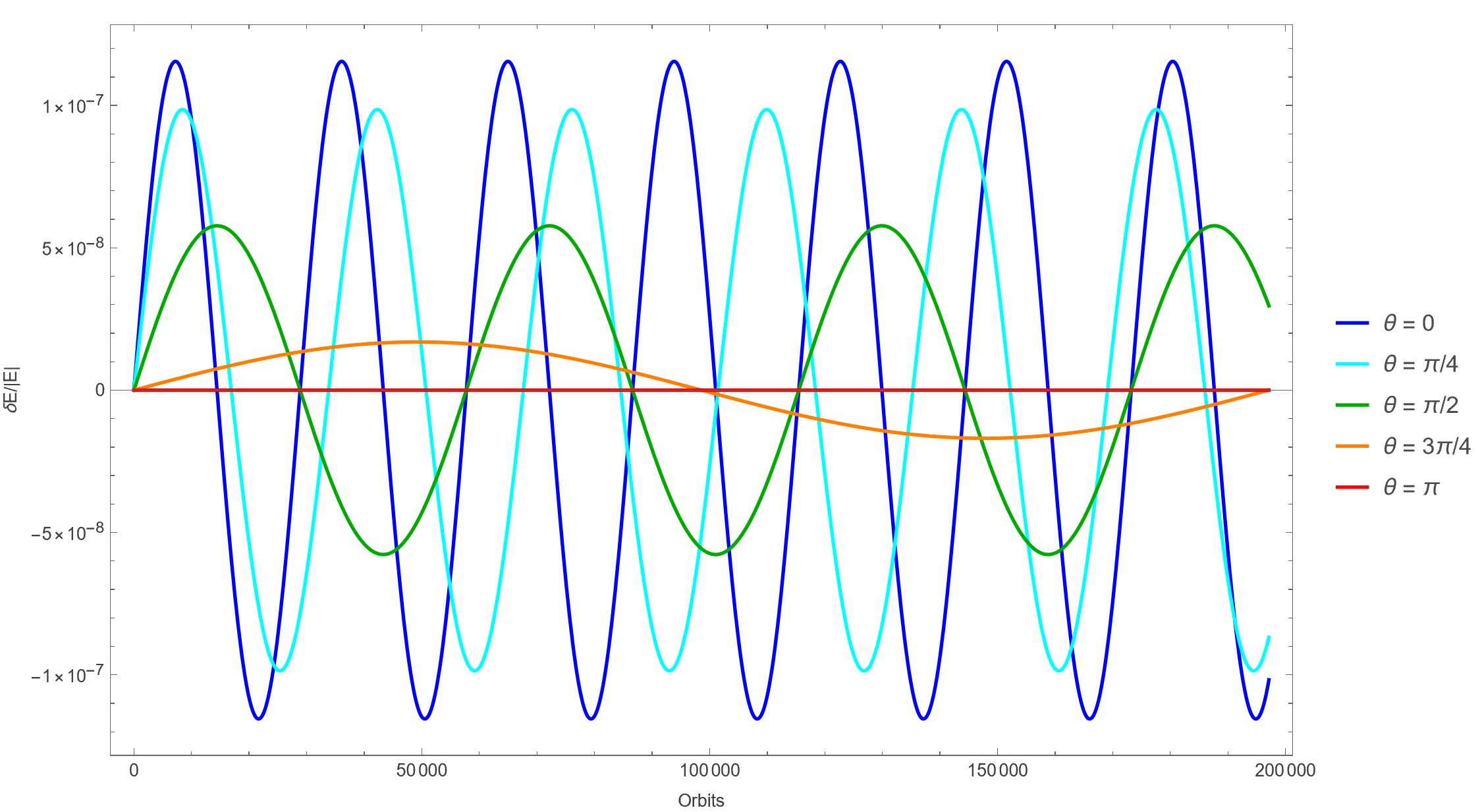}}\hfill
	\subfloat[Mean motion]{\label{theta_c}\includegraphics[height=0.3\textheight]{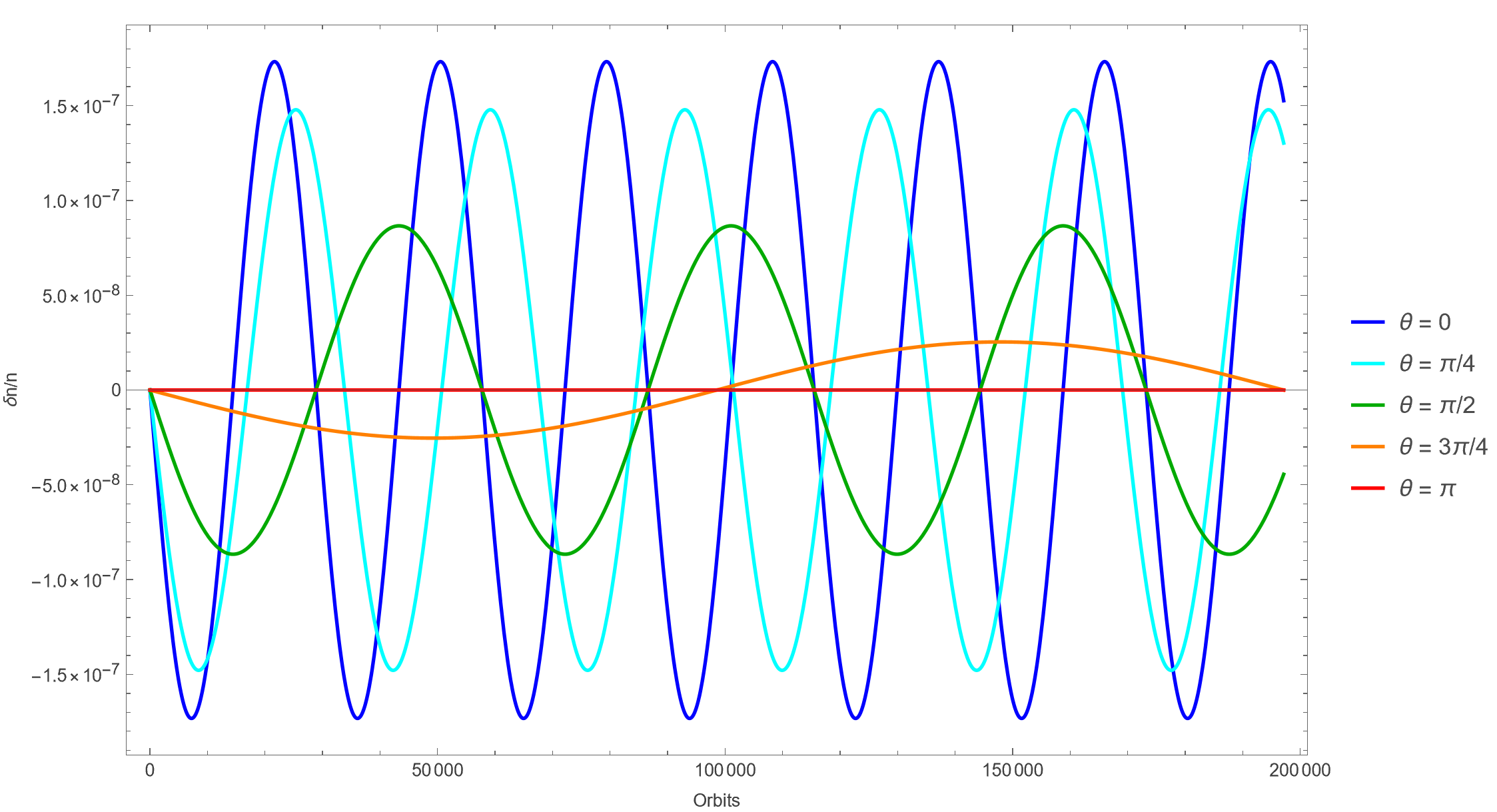}}
	\caption{Relative change in semi-major axis, specific energy and mean motion for different values of GW-particle inclination for $\omega_{\text{GW}} = 2n_{0}$. Initial conditions are $h_{0}=10^{-10}$, $a=1$, $\theta=0$ and $\Delta=-0.01$}
	\label{fig:theta}
\end{figure}

The inclination angle between the GW and the particle's orbital plane, $\theta$, has an effect on the energy exchange. Appearing in Equation (\ref{C+}) and (\ref{C-}), $\theta$ modulates the amplitude coefficients $C_{+}$ and $C_{-}$. This is illustrated in Figure \ref{fig:theta} where we orbit average to negate the effects of $\Phi$. It is clear that the maximum amount of energy is exchanged when the GW is incident directly along the $z$-axis ($\theta = 0, |C_-| = h_0$). This was discussed in \S\ref{R&P} alongside Figure \ref{fig:hel} and had been made explicit here. As $\theta$ sweeps through from $0$ to $\pi/2$, towards an edge-on configuration, the GW contributes decreasing amounts of energy to the system while also increasing the libration cycle. This is consistent with the reduction in the resonant coupling coefficient ($|C_{+}| = |C_{-}| = 0.25h_{0}$). 

For $\theta=3\pi/4$, the resonant response is substantially weaker, with just one libration cycle occurring over the entire integration interval. At $\theta=\pi$, the resonant coefficient vanishes identically as $C_-=0$, and no resonant libration occurs in the orbit-averaged model. The angular dependence is therefore asymmetric about $\theta=\pi/2$, reflecting the helicity-dependent coupling between the circularly polarized GW and the orbital motion.

These results are consistent with the analytical small-amplitude libration frequency of Equation \eqref{Omega_lib} and the corresponding dependence of the libration period on incidence angle. They further demonstrate that the approximately linear energy evolution observed over short timescales does not represent indefinite secular growth. Instead, the resonant energy exchange reverses periodically, remaining bounded within the libration region.

The calculation assumes a fixed orbital plane and initially circular Keplerian motion. It therefore describes the angular dependence of the reduced resonant dynamics rather than the complete evolution of all osculating orbital elements.








\subsection{Evolution of the Orbital Elements}
\label{Elements}

The preceding analytical treatment assumes initially circular Keplerian motion in a fixed orbital plane. However, the tidal acceleration induced by the incident GW generally possesses radial, transverse and normal components, which can perturb the eccentricity and orientation of the orbit in addition to its semi-major axis \cite{MaCaHu1981}. To assess the validity of the reduced circular-orbit approximation, we therefore consider the evolution of all six osculating orbital elements using Gauss' variational equations (GVEs) which describe the response of a Keplerian orbit to an arbitrary perturbing acceleration \cite{Iorio2025}.

Since the conventional orbital elements become singular at zero eccentricity and inclination, the numerical integration is performed using non-singular modified equinoctial elements \cite{WaIrOw1985}. The GW perturbation is evaluated using the instantaneous osculating position and velocity, allowing the eccentricity, inclination, longitude of the ascending node and orientation of the eccentricity vector to evolve without imposing a fixed orbital plane. The corresponding equations and coordinate transformations are presented in Appendix \ref{Inc}.

\begin{figure}
	\centering
	\includegraphics[width=\textwidth]{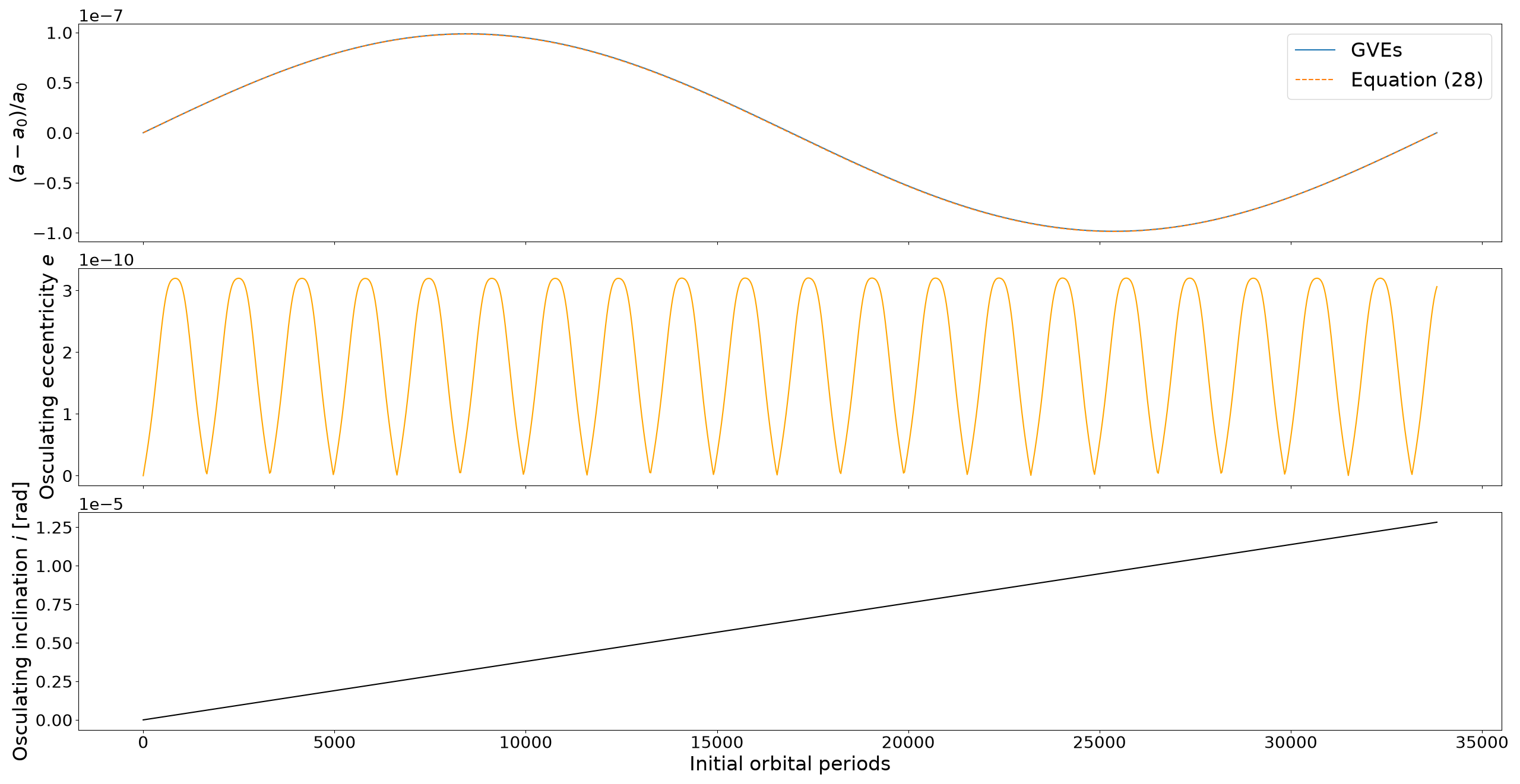}
	\caption{Osculating orbital elements evolved for one libration period using $h_0=10^{-10}$, $\theta=\pi/4$, $\omega_{\mathrm{GW}}=2n_0$ and $\Delta(0)=-0.01$.}
	\label{fig:elements}
\end{figure}

The six-element equations were integrated over one libration period with the results presented in Figure \ref{fig:elements}. We observe that the eccentricity undergoes small oscillations, reaching a maximum of $3.20\times10^{-10}$ with a period of approximately $T_{\rm lib}/20.5$. On the other hand, the inclination increases from an initial value of 0 to approximately $1.28\times10^{-5}$ radians. The latter demonstrates that the orbital plane is not strictly stationary, even in the weak-forcing regime.

Despite these perturbations, the semi-major axis evolution remains in close agreement with the reduced circular orbit model. The GPE integration predicts a maximum fractional semi-major axis excursion of approximately $9.88\times10^{-8}$, compared with $9.86\times10^{-8}$ obtained from the orbit-averaged pendulum equations. The maximum point-wise discrepancy over the sampled libration interval is $4.21\times10^{-10}$, corresponding to approximately 0.4\% of the principal resonant excursion. The effective GW incidence angle changes by less than $2.1\times10^{-8}$ radians in this configuration.

These results indicate that the circular, fixed-plane approximation accurately reproduces the leading resonant semi-major axis dynamics for the investigated large-strain initial conditions, although eccentricity and inclination are not exactly conserved. In particular, the small cumulative inclination change does not significantly modify the resonant energy exchange over the simulated interval.

This conclusion should not be interpreted as a general conservation result for the remaining orbital elements. The magnitude and character of their perturbations depend on the GW polarization, incidence angle, initial orbital phase and strain amplitude. The reduced pendulum model is therefore best regarded as a leading-order approximation whose applicability must be assessed for the orbital configuration under consideration.

\section{Conclusion}

We have derived an analytical expression for the rate of change of specific energy of a circular binary due to the continuous passing of a GW through the system. The GW imparts positive energy into the system leading to an increase in an orbiting particle's semi-major axis and subsequent decrease in mean motion. This effect is influenced by the GWs angle of incidence which dictates the strength of the changes. 

The corresponding dynamics are described by a driven pendulum equation with a stable equilibrium at $\Delta=0$ and libration frequency $\omega_{\mathrm{lib}}=\omega_{\mathrm{GW}}\sqrt{3|C_-|}$. For the adopted polarization handedness, the resonant coupling is strongest at $\theta=0$, decreases with incidence angle, and vanishes at $\theta=\pi$. Reversing the wave's handedness reverses this angular preference.

The validity of the initially circular, fixed-plane approximation has additionally been assessed by integrating the GVEs. Eccentricity and inclination are not strictly conserved. Nevertheless, for the illustrative strong-forcing configuration considered, the principal semi-major axis evolution agrees closely with the reduced pendulum model over one libration period, with a maximum sampled discrepancy of approximately 0.4\% of the resonant excursion.

The exchange of energy is dependent upon two oscillation modes (fast and slow) which restrict the type of astrophysical systems which are of interest as only those with very small orbital periods like pulsars will produce librating orbits over potentially observational timescales. Resonant interactions between systems with longer orbital periods is dynamically irrelevant as from the current analysis, any secular effect will not be observable. 

\medskip
\textbf{Acknowledgments} \par 
The authors would like to thank Charlie Sharpe, discussions with whom allowed us to apply their formalism to our work. Thank you also to Helen Christodoulidi for providing feedback on early manuscript drafts as well as to the anonymous reviewers of this manuscript for their constructive comments. 

\section{References}

\printbibliography[heading=none]

\appendix

\section{Rate of Change of Energy}
\label{Energy}

What follows is the derivation of equations from the main text. Starting from the particles frame where

\begin{align}
	X(t) = a\cos(nt) && Y(t) = a\sin(nt)
\end{align}

\begin{align}
	\dot{X} = -an\sin(nt) && \dot{Y} = an\cos(nt)
\end{align}

In the frame of the GW

\begin{align}
	x = [X\cos(\phi) + Y\sin(\phi)]\cos(\theta) && y = -X\sin(\phi) + Y\cos(\phi)
\end{align}

By defining $\psi = (nt-\phi)$ and using the identity $\cos(A-B) = \cos(A)\cos(B) + \sin(A)\sin(B)$, we achieve

\begin{align}
	x = a\cos(\psi)\cos(\theta) && y = a\sin(\psi)
	\label{xy}
\end{align}

\begin{align}
	\dot{x} = -an\sin(\psi)\cos(\theta) && \dot{y} = an\cos(\psi) && \dot{\psi} = n
	\label{dot_xy}
\end{align}

Kepler's third law tell us the orbital energy of the system in the center-of-mass frame is 

\begin{equation}
	\mathcal{E} = \frac{1}{2}\mu_r|\dot r^2| - \frac{\mu_r\mu_g}{r} = -\frac{\mu_r\mu_g}{2a}
	\label{energy}
\end{equation}

where $\mu_r$ is the systems reduced mass which in systems where the mass of the central body is much larger than that of the orbiting body ($M\gg m$) becomes $\mu_r = \frac{Mm}{M + m} = m$. $\mu_g$ is the gravitational parameter which is also simplified in our system, $\mu_g = GM = M$.

The acceleration is independent of the test-particle mass, but the total energy transferred by an incident GW is proportional to $\mu_r$ such that

\begin{equation}
	\dot{\mathcal{E}} = \frac{\mu_r}{2}\left[\ddot{h}_{+}(x\dot{x} - y\dot{y}) + \ddot{h}_{\times}(x\dot{y} + y\dot{x})\right]
	\label{e_dot}
\end{equation}

When working with the specific energy ($E=\mathcal{E}/\mu_r$) we see that Equations \eqref{energy} and \eqref{e_dot} become

\begin{align}
	E = -\frac{\mu_g}{2a}, \quad \dot{E} = \frac{1}{2}\left[\ddot{h}_{+}(x\dot{x} - y\dot{y}) + \ddot{h}_{\times}(x\dot{y} + y\dot{x})\right]
	\label{E_dot}
\end{align}

The geometric terms are

\begin{align}
	x\dot{x} - y\dot{y} = &&(-an\sin(\psi)\cos(\theta))(a\cos(\psi)\cos(\theta)) - (an\cos(\psi))(a\sin(\psi)) \\
	&& = -a^{2}n\sin(\psi)\cos(\psi)(\cos^{2}(\theta) + 1) \\
	&& = -\frac{1}{2}a^{2}n(\cos^{2}(\theta) + 1)\sin(2\psi)
	\label{xx-yy}
\end{align}

where we have used $\sin(2A) = 2\sin(A)\cos(A)$. 

The next term is

\begin{align}
	x\dot{y} + y\dot{x} = &&(a\cos(\psi)\cos(\theta))(an\cos(\psi)) + (a\sin(\psi))(-an\sin(\psi)\cos(\theta)) \\
	&& = a^{2}n\cos(\theta)(\cos^{2}(\psi) - \sin^{2}(\psi)) \\
	&& = a^{2}n\cos(\theta)\cos(2\psi)
	\label{xy+yx}
\end{align}

where the identity $\cos(2A) = \cos^{2}(A) - \sin^{2}(A)$ has been employed.

For a circularly polarized GW where

\begin{align}
	h_{+} = h_{0}\cos(\omega_{\text{GW}}t -kz) && h_{\times} = h_{0}\sin(\omega_{\text{GW}}t -kz)
\end{align}

\begin{align}
	\ddot{h}_{+} = -\omega_{\text{GW}}^2h_{0}\cos(\omega_{\text{GW}}t) && \ddot{h}_{\times} = -\omega_{\text{GW}}^2h_{0}\sin(\omega_{\text{GW}}t)
	\label{ddot_h}
\end{align}

Substituting Equations (\ref{xx-yy}), (\ref{xy+yx}) and (\ref{ddot_h}) into Equation (\ref{E_dot}) requires the use of $\sin(A)\cos(B) = 1/2[\sin(A+B)+\sin(A-B)]$ which produces

\begin{align}
	\dot{E} = \frac{a^{2}n\omega_{\text{GW}}^2}{2}\Big[\frac{1}{2}h_{0}(1+\cos^{2}(\theta))\cos(\omega_{\text{GW}}t)\sin(2\psi) - h_{0}\cos(\theta)\sin(\omega_{\text{GW}}t)\cos(2\psi)\Big] \\
	= \frac{a^2n\omega_{\text{GW}}^2}{2}[C_{+}\sin((\omega_{\text{GW}} + 2n)t-2\phi) + C_{-}\sin((\omega_{\text{GW}} -2n)t + 2\phi)]
\end{align}

where we define two amplitude coefficients

\begin{align}
	C_{+} = \frac{1}{4}h_{0}(1-\cos(\theta))^2 && C_{-} = -\frac{1}{4}h_{0}(1+\cos(\theta))^2
	\label{C+}
\end{align}

\subsection{Polarization Modes}
\label{Polarisations}

Whereas the main text only considered circularly polarized GWs, detailed here is the construction of pendulum equations for the pure $+$, pure $\times$ and mixed polarization GW cases.

\subsection{Plus Polarization}
\label{+}

For a purely $+$-polarized GW the amplitude of the $\times$-polarization is zero ($h_{\times} = 0$) and the $+$-polarization amplitude takes the form

\begin{equation}
	h_{+} = h_{0,+}\cos(\omega_{\rm GW}t)
	\label{h+}
\end{equation}

and 

\begin{equation}
	\ddot{h}_{+} = -\omega_{\rm GW}^{2}h_{0,+}\cos(\omega_{\rm GW}t)
	\label{ddot_h+}
\end{equation}

Starting from the full rate of change of energy given by Equation (\ref{E_dot}), we can rewrite it using Equation (\ref{xy}), (\ref{dot_xy}) and (\ref{ddot_h+}) such that

\begin{equation}
	\begin{split}
		\dot{E}_{+} = \frac{a^{2}n\omega_{\rm GW}^{2}h_{0,+}}{4}(1+\cos^{2}(\theta))\cos(\omega_{\text{GW}}t)\sin(2\psi) \\
		=\frac{a^{2}n\omega_{\rm GW}^{2}h_{0,+}}{8}(1+\cos^{2}(\theta))\sin(\Phi_{+}) - \sin(\Delta_{+})]
	\end{split}
	\label{dot_E+}
\end{equation}

where we have used $\cos(A)\sin(B)=\frac{1}{2}[\sin(B+A)+\sin(B-A)]$ and defined fast ($\Phi_{+}$) and slow ($\Delta_{+}$) resonant terms

\begin{equation}
	\Phi_{+} = \omega_{\rm GW}t - 2\psi
\end{equation}

\begin{equation}
	\Delta_{+} = \omega_{\rm GW}t + 2\psi
\end{equation}

allows us to write Equation (\ref{dot_E+}) as

\begin{equation}
	\dot{E}_{+} = -\frac{a^{2}n\omega_{\rm GW}^{2}}{2}\Big[P_{+}\sin(\Phi_{+}) + P_{-}\sin(\Delta_{+})\Big]
	\label{dot_E_+}
\end{equation}

where 

\begin{equation}
	P_{\pm} = \pm\frac{h_{0,+}}{4}(1 + \cos^{2}(\theta))
	\label{P}
\end{equation}

Averaging out the fast term leaves

\begin{equation}
	\langle\dot{E}_{+}\rangle = -\frac{a^{2}n\omega_{\text{GW}}^{2}h_{0,+}}{8}(1+\cos^{2}(\theta))\sin(\Delta_{+}).
\end{equation}

By considering the second time differential of $\Delta_{+}$, we can produce an equation which described the motion of a pendulum. Following the same derivation as Equations (21)--(25),

\begin{equation}
	\ddot{\Delta}_{+}+\frac{3}{4}h_{0,+}(1+\cos^{2}(\theta))\omega_{\text{GW}}^{2}\sin(\Delta_{+})=0
	\label{Pendulum_+}
\end{equation}

Thus the resonant dynamics are governed by a pendulum equation whose effective amplitude is

\begin{equation}
	A_{+}=\frac{3}{4}h_{0,+}\omega_{\text{GW}}^{2}(1+\cos^{2}(\theta))
\end{equation}

The corresponding libration frequency is

\begin{equation}
	\omega_{\rm lib,+}	= \omega_{\text{GW}}\sqrt{\frac{3}{4}h_{0,+}(1+\cos(\theta))^{2}}
\end{equation}

\subsection{Cross Polarization}
\label{x}

Similar to \S\ref{+}, we start from Equation (\ref{E_dot}) and apply a $\times$-polarized wave where $h_{+} = 0$ and 

\begin{align}
	h_{\times} = h_{0,\times}\cos(\omega_{\rm GW}t), && \ddot{h}_{\times} = -\omega_{\rm GW}^{2}h_{0,\times}\cos(\omega_{\rm GW}t)
	\label{hx}
\end{align}

Equation (\ref{E_dot}) becomes

\begin{equation}
	\dot{E}_{\times}=-\frac{a^{2}n\omega_{\rm GW}^{2}h_{0,\times}}{2}\cos(\theta)\cos(\omega_{\text{GW}}t)\cos(2\psi)
\end{equation}

Using the identity $\cos(A)\cos(B)=\frac{1}{2}\left[\cos(A+B)+\cos(A-B)\right]$ gives

\begin{align}
	\dot{E}_{\times}=-\frac{a^{2}n\omega_{\text{GW}}^{2}h_{0,\times}}{4}\cos(\theta)[\cos(\Phi_{\times})+\cos(\Delta_{\times})]
\end{align}

after defining similar fast ($\Phi_{\times}$) and slow ($\Delta_{\times}$) resonance terms. 

Averaging out the fast term leaves

\begin{equation}
	\langle\dot{E}_{\times}\rangle=-\frac{a^{2}n\omega_{\text{GW}}^{2}h_{0,\times}}{4}\cos(\theta)\cos(\Delta_\times)
\end{equation}

Proceeding as before yields

\begin{equation}
	\ddot{\Delta}_{\times}+\frac{3}{2}\omega_{\text{GW}}^{2}h_{0, \times}\cos(\theta)\cos(\Delta_{\times})=0
	\label{ddot_Delta_x}
\end{equation}

Introducing the relationship

\begin{equation}
	\Psi_{\times}=\Delta_\times+\frac{\pi}{2}
\end{equation}

and using

\begin{equation}
	\cos(\Delta_\times) = \sin(\Psi_\times)
\end{equation}

we can transform Equation (\ref{ddot_Delta_x}) into

\begin{equation}
	\ddot{\Psi}_{\times}+\frac{3}{2}\omega_{\text{GW}}^{2}h_{0,\times}\cos(\theta)\sin(\Psi_{\times})=0
	\label{Pendulum_x}
\end{equation}

For $h_{0,\times}\cos(\theta)<0$ the stable equilibrium is shifted by $\pi$ instead. In either case, if the coupling is nonzero, the small-amplitude libration frequency about the stable equilibrium is

\begin{equation}
	\omega_{\rm lib,\times}=\omega_{\text{GW}}\sqrt{\frac{3}{2}|h_{0,\times}\cos(\theta)|}
\end{equation}

\subsection{Mixed Polarization}
\label{Mixed}

Consider the general polarization state

\begin{equation}
	h_{+}=h_{0, +}\cos(\omega_{\rm GW}t)
\end{equation}

\begin{equation}
	h_{\times}=h_{0, \times}\cos(\omega_{\rm GW}t+\delta)
\end{equation}

where $\delta$ is the phase difference between the two polarization modes.

After substitution into Equation (\ref{E_dot}) and averaging over the fast oscillation, the resonant energy exchange becomes

\begin{align}
	\langle\dot{E}\rangle_{\rm res}=-\frac{1}{8}a^{2}n\omega_{\text{GW}}^{2}h_{0, +}	(1+\cos^{2}(\theta))\sin(\Delta)-\frac{1}{2}a^{2}n\omega_{\text{GW}}^{2}h_{0, \times}\cos(\theta)\cos(\Delta+\delta)
\end{align}

Expanding $\cos(\Delta+\delta) = \cos(\Delta)\cos(\delta) - \sin(\Delta)\sin(\delta)$ gives

\begin{equation}
	\langle\dot{E}\rangle_{\rm res}=K[S\sin(\Delta)+C\cos(\Delta)]
\end{equation}

where

\begin{equation}
	K=\frac{1}{2}a^{2}n\omega_{\text{GW}}^{2}
\end{equation}

\begin{equation}
	S=-\frac{1}{4}h_{0, +}(1+\cos^{2}\theta)+\frac{1}{2}h_{0, \times}\cos(\theta)\sin(\delta)
\end{equation}

and

\begin{equation}
	C=-\frac{1}{2}h_{0, \times}\cos(\theta)\cos(\delta)
\end{equation}

Defining

\begin{equation}
	R=\sqrt{S^{2}+C^{2}}, \quad \beta=\atantwo(C,S), \quad S = R\cos(\beta), \quad C=R\sin(\beta)
\end{equation}

allows the resonant energy exchange to be written compactly as

\begin{equation}
	\langle\dot{E}\rangle_{\rm res}=KR\sin(\Delta+\beta)
\end{equation}

Repeating the pendulum equation derivation gives

\begin{equation}
	\ddot\Delta-3\omega_{\text{GW}}^{2}R\sin(\Delta+\beta)=0
\end{equation}

Introducing a shifted phase

\begin{equation}
	\Gamma=\Delta+\beta-\pi
\end{equation}

yields the final pendulum equation

\begin{equation}
	\ddot\Gamma+3\omega_{\rm GW}^{2}R\sin(\Gamma)=0
	\label{Pendulum_Mixed}
\end{equation}



The pendulum-like resonance trap produced by a mixed polarization GW is bounded by an effective amplitude, $C$ that is dependent on the relative phase difference between the two modes. 

Equation (\ref{Pendulum_Mixed}) reduces to the equation for a circularly polarized GW when $h_{+} = h_{\times} = h_{0}$ and $\delta = -\frac{\pi}{2}$. Circular polarization maximizes the coupling for certain inclinations because contributions from the $+$ and $\times$-polarization's combine coherently, whereas pure $+$ and pure $\times$ GWs excite the resonance through only one quadrupolar channel.

\section{Gauss' Variational Equations (GVEs)}
\label{GVEs}

The preceding energy and pendulum calculations assume an initially circular osculating orbit and a fixed orbital plane. They do not establish that eccentricity and inclination remain negligible over a full libration period. For a fully self-consistent test, the tidal acceleration must be resolved into radial ($A_R$), transverse ($A_T$), and normal ($A_N$) components and the corresponding orbital elements evolved. For the initially circular, prograde orbit the argument of periapsis ($\omega$) and longitude of the ascending node ($\Omega$) are not separately defined at $e=i=0$. To integrate the response without introducing artificial non-zero initial eccentricity or inclination, we adopt the modified equinoctial elements 

\begin{equation}
	(p,f,g,h,k,L)=\bigl(a(1-e^2),\,e\cos(\Omega+\omega),\,
	e\sin(\Omega+\omega),\,\tan(i/2)\cos(\Omega),\,
	\tan(i/2)\sin(\Omega),\,\Omega+\omega+\nu\bigr),
\end{equation}

where $\nu$ is the true anomaly and the test-particle mass has been divided out of the orbital acceleration.

Let
\begin{equation}
	w=1+f\cos(L)+g\sin(L), \quad s^2=1+h^2+k^2, \quad \zeta=h\sin(L)-k\cos(L)
\end{equation}

Then for the non-Keplerian tidal acceleration $(A_R,A_T,A_N)$, the six GVEs are

\begin{align}
	\dot p &= 2\frac{p}{w}\sqrt{\frac p{\mu_g}}A_T,\\
	\dot f &=\sqrt{\frac p{\mu_g}}\left[
	\sin L\,A_R+\frac{(w+1)\cos L+f}{w}A_T
	-\frac{g\zeta}{w}A_N\right],\\
	\dot g &=\sqrt{\frac p{\mu_g}}\left[
	-\cos L\,A_R+\frac{(w+1)\sin L+g}{w}A_T
	+\frac{f\zeta}{w}A_N\right],\\
	\dot h &=\sqrt{\frac p{\mu_g}}\frac{s^2\cos L}{2w}A_N,\\
	\dot k &=\sqrt{\frac p{\mu_g}}\frac{s^2\sin L}{2w}A_N,\\
	\dot L &=\sqrt{\mu_gp}\left(\frac{w}{p}\right)^2
	+\sqrt{\frac p{\mu_g}}\frac{\zeta}{w}A_N.
\end{align}

At every integration step, we reconstruct the osculating position and velocity from the equinoctial basis

\begin{align}
	\mathbf F &=s^{-2}(1+h^2-k^2,\,2hk,\,-2k),\\
	\mathbf G &=s^{-2}(2hk,\,1-h^2+k^2,\,2h),\\
	\mathbf r&=\frac p w(\cos L\,\mathbf F+\sin L\,\mathbf G),\\
	\mathbf v&=\sqrt{\frac{\mu_g}p}
	\bigl[-(g+\sin L)\mathbf F+(f+\cos L)\mathbf G\bigr]
	\label{GVE}
\end{align}

With $\hat{\mathbf R}=\mathbf r/r$, $\hat{\mathbf N}=(\mathbf r\times\mathbf v)/|\mathbf r\times\mathbf v|$, and $\hat{\mathbf T}=\hat{\mathbf N}\times\hat{\mathbf R}$, the tidal projections are $A_j=\mathbf A_{\mathrm{GW}}\cdot\hat{\mathbf j}$.

The polarization basis is $\mathbf P=(\cos(\theta),0,-\sin(\theta))$, $\mathbf Q=(0,1,0)$ and the tidal acceleration can be determined from the instantaneous osculating position via

\begin{align}
	\mathbf A_{\mathrm{GW}}=-\frac{h_0\omega_{\mathrm{GW}}^2}{2}
	\bigl\{\cos(\omega_{\mathrm{GW}}t+\zeta_0)
	[(\mathbf P\cdot\mathbf r)\mathbf P-(\mathbf Q\cdot\mathbf r)\mathbf Q]+\sin(\omega_{\mathrm{GW}}t+\zeta_0)
	[(\mathbf Q\cdot\mathbf r)\mathbf P+(\mathbf P\cdot\mathbf r)\mathbf Q]\bigr\}
\end{align}

Then $A_j = \textbf{A}_{\rm GW}\cdot\hat{\textbf{j}}$ with $j = R, T, N$. The initial values are $p=a_0$, $f=g=h=k=0$, $L=0$, so that an exactly circular initial orbit is represented without coordinate singularities. The node and pericentre orientation are recoverable wherever $i$ and $e$ are nonzero but are undefined at the corresponding zero-element limit. The classical elements are reconstructed wherever defined as $a=p/(1-f^2-g^2)$, $e=\sqrt{f^2+g^2}$, $i=2\arctan\sqrt{h^2+k^2}$, $\Omega=\mathrm{atan2}(k,h)$, and $\varpi=\mathrm{atan2}(g,f)$; then $\omega=\varpi-\Omega$ and the mean anomaly follows from the usual eccentric-anomaly conversion. At isolated $e=0$ or $i=0$ values, these angles have no unique numerical meaning; the regularized elements
remain well defined.

\subsection{Inclination Evolution}
\label{Inc}

Take an initially circular reference orbit $\textbf{r} = a(\cos(u), \sin(u), 0)$ with $u=n_ot + u_0$ and write the GW phase as $\psi_{\rm GW} = \omega_{\rm GW} + \xi_0$. The normal component of the GW acceleration on an initially circular orbit is

\begin{equation}
	\textbf{A}_{N} = \frac{h_0\omega_{\rm GW}^2a\sin(\theta)}{4}[(1+\cos(\theta))\cos(\psi-u) - (1-\cos(\theta))\cos(\psi+u)]
	\label{A_N}
\end{equation}

At the $\omega_{\rm GW}=2n_0$ resonance, the trigonometric arguments in Equation \eqref{A_N} contain frequencies $n_0$ and $3n_0$. However, the inclination variables in Equation \eqref{GVE} contain additional factors of $\cos(u)$ and $\sin(u)$, which when multiplied through Equation \eqref{A_N} produces non-oscillatory terms. 

By writing the slowly evolving resonant phase as $\Delta - \psi - 2u$ and averaging over the orbital motion yields the following to leading order in inclination and eccentricity

\begin{equation}
	\left\langle\dot h\right\rangle = \frac{h_0n_0}{4}\sin(\theta)(1+\cos(\theta))\cos(\Delta)
\end{equation}

\begin{equation}
	\left\langle\dot k\right\rangle=-\frac{h_0n_0}{4}\sin(\theta)(1+\cos(\theta))\sin(\Delta)
\end{equation}

Near the stable equilibrium, $|\Delta|\ll1$, leading to $\cos(\Delta)\simeq1$. Therefore, over a period during which the small-angle approximation remains valid and we have

\begin{equation}
	h(t)\simeq\frac{h_0n_0}{4}\sin(\theta)(1+\cos(\theta))t
\end{equation}

For small inclination, $i\simeq2\sqrt{h^2+k^2}$, giving the approximate leading drift as

\begin{equation}
	i(t)\simeq\frac{h_0n_0t}{2}\sin(\theta)(1+\cos(\theta))
	\label{i}
\end{equation}

Using the libration period, $T_{\rm lib}=\frac{2\pi}{n_0\sqrt{3h_0}(1+\cos\theta)}$, Equation \eqref{i} predicts 

\begin{equation}
	i(T_{\rm lib})\simeq\pi\sqrt{\frac{h_0}{3}}\sin(\theta)	
\end{equation}

for a trajectory initially close to the stable resonant equilibrium and remaining within the strong GW strain approximation. Using illustrative parameters $h_0=10^{-10}$ and $\theta=\pi/4$ produces $i(T_{\rm lib})\simeq1.283\times10^{-5}\ {\rm rad}$. This agrees with the approximately $1.28\times10^{-5}$ rad inclination change found in the GVE calculation within the main text. The agreement demonstrates that the inclination change can accumulate to order $\sqrt{h_0}$ over a libration timescale, even though its instantaneous rate is of order $h_0n_0$.
	
\end{document}